\documentclass{article}

\PassOptionsToPackage{numbers, sort&compress}{natbib}
\usepackage[main, final]{neurips_2026}
\usepackage[utf8]{inputenc} % allow utf-8 input
\usepackage[T1]{fontenc}    % use 8-bit T1 fonts
\usepackage{hyperref}       % hyperlinks
\usepackage{url}            % simple URL typesetting
\usepackage{booktabs}       % professional-quality tables
\usepackage{amsfonts}       % blackboard math symbols
\usepackage{nicefrac}       % compact symbols for 1/2, etc.
\usepackage{microtype}      % microtypography
\usepackage{xcolor}         % colors
\usepackage{multirow}
\usepackage{graphicx}
\usepackage{amsmath}
\usepackage{amssymb}
\usepackage{amsthm}
\usepackage{dsfont}
\usepackage{float}

\title{SENSE: Semantic Neural Speech Synthesis from \\Brain Dynamics via Spatial Graph Encoding}

\author{%
  Jisoo Park$^{1}$\thanks{Work done during a research visit to the University of Birmingham.} \quad
  Seonghak Lee$^{2}$ \quad
  Hyojin Park$^{3}$ \quad
  Junseok Kwon$^{1,2}$ \\
  $^{1}$Department of Artificial Intelligence, Chung-Ang University, Korea \\
  $^{2}$Department of Computer Science and Engineering, Chung-Ang University, Korea \\
  $^{3}$School of Psychology, Centre for Human Brain Health (CHBH), University of Birmingham, UK\\
  \texttt{susiehome@cau.ac.kr}
}

\begin{document}

\maketitle

\begin{abstract}
 Reconstructing speech from non-invasive brain signals offers a promising pathway for restoring communication in individuals who are cognitively intact but unable to speak. Existing EEG-to-speech approaches formulate this task as \emph{acoustic reconstruction}, optimizing waveform fidelity while ignoring whether the generated speech preserves high-level semantic content. In this work, we revisit this formulation and argue that EEG signals carry not only acoustic but also semantic information. We identify two key limitations of prior methods: (1) the neglect of spatial relationships between EEG electrodes, and (2) the failure to exploit the semantic structure of the N400 paradigm, where congruent and incongruent trials reflect distinct semantic processing. We propose \textbf{SENSE} (\textbf{S}emantic-\textbf{E}EG \textbf{N}eural \textbf{S}peech Synth\textbf{E}sis), which combines a graph-based EEG encoder over electrode geometry with EEG Semantic Conditioning (ESC), aligning EEG to a pretrained semantic space using only congruent trials. On the N400 dataset, SENSE consistently outperforms prior methods on both acoustic and semantic metrics, and model-internal channel attribution suggests distributed reliance on auditory, sensorimotor, and centro-parietal regions, consistent with known speech-perception neuroscience. In the unseen-subject setting, SENSE trained on only two subjects already surpasses the strongest baseline trained on all eighteen subjects in word error rate, and matches it on acoustic metrics with as few as eight subjects. \setcounter{footnote}{0}\footnote{Code and audio samples are available at \url{https://jisoo-o.github.io/website/projects/SENSE/}.}
\end{abstract}
       %\vspace{-2mm}
\section{Introduction}
\label{sec:intro}
       %\vspace{-3mm}
Restoring speech communication from brain activity is a central goal of brain-computer interface (BCI) research, with direct implications for individuals who are cognitively intact but unable to speak due to conditions such as amyotrophic lateral sclerosis (ALS) or locked-in syndrome~\cite{willett2023high,metzger2023high}. Beyond speech, neural decoding has enabled motor intention decoding~\cite{mcfarland2011brain,simeral2011neural}, visual stimulus reconstruction~\cite{takagi2023high,lan2023seeing}, and decoding of high-level semantic representations~\cite{tang2023semantic,pereira2018toward}. State-of-the-art results, however, largely rely on invasive or high-fidelity modalities such as electrocorticography (ECoG) and functional MRI (fMRI), which are limited by surgical requirements or low temporal resolution~\cite{ullsperger2010simultaneous}. Among non-invasive alternatives, electroencephalography (EEG) is particularly attractive due to its portability, low cost, and millisecond-level temporal resolution~\cite{luck2014,abiri2019comprehensive,roy2019deep}.

%Neural decoding has enabled a broad spectrum of applications beyond speech, including motor intention decoding~\cite{mcfarland2011brain,schalk2004bci2000,simeral2011neural}, text generation from neural signals~\cite{willett2021high,duan2023dewave,tang2023semantic}, reconstruction of visual stimuli from brain activity~\cite{shen2019deep,horikawa2017hierarchical,takagi2023high,lan2023seeing}, and decoding of high-level semantic representations~\cite{huth2016natural,pereira2018toward,tang2023semantic}. Despite this progress, state-of-the-art performance largely depends on invasive or high-fidelity modalities such as electrocorticography (ECoG) and functional MRI (fMRI), which are limited by surgical requirements or low temporal resolution~\cite{ullsperger2010simultaneous}.

%Among non-invasive alternatives, electroencephalography (EEG) is particularly attractive due to its portability, low cost, and millisecond-level temporal resolution~\cite{luck2014,abiri2019comprehensive,roy2019deep}. However, EEG signals are inherently noisy, susceptible to artifacts, and only indirectly reflect underlying neural sources, making representation learning and source localization difficult~\cite{roy2019deep,nunez2006electric,chen2022toward}. These challenges are further exacerbated in the small-data regime typical of EEG studies~\cite{chen2022toward}.

\begin{figure}
    \centering
    \includegraphics[width=1.0\textwidth]{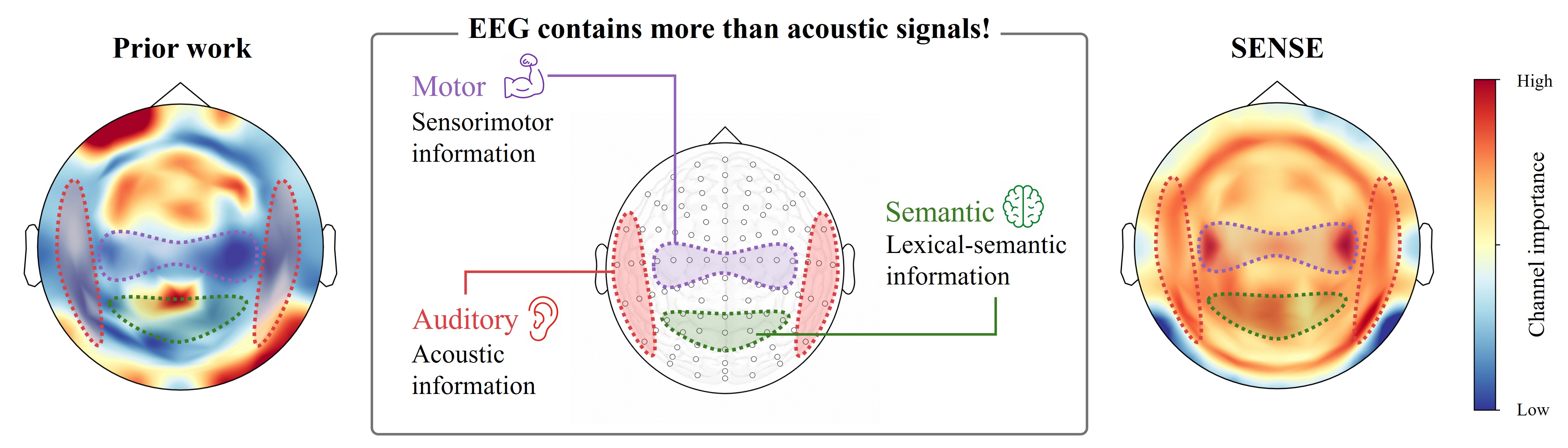}
    %\vspace{-2mm}
    \caption{\textbf{Motivation for the proposed SENSE framework in EEG-to-Speech Generation.} EEG during speech perception carries acoustic, sensorimotor, and lexical-semantic information distributed across the scalp (center). Prior work~\cite{lee2025phoneme} captures channel patterns without clear anatomical structure (left); SENSE recovers attribution aligned with all three functional regions (right) via  \emph{EEG Semantic Conditioning} and \emph{Graph-based EEG Encoding}. }%Channel-importance maps are obtained from gradient-based attribution and normalized per model.}
    \label{fig:teaser}
        %\vspace{-3mm}
\end{figure}

Despite EEG's well-known limitations of noise, artifact susceptibility, and indirect reflection of neural sources~\cite{roy2019deep,nunez2006electric,chen2022toward}, recent work has shown that speech can be reconstructed from EEG in \emph{perceived speech} settings, where neural activity recorded during passive listening is paired with the speech signal. Early approaches predict acoustic features and then synthesize speech from them~\cite{krishna2020speech}, while more recent methods directly generate waveforms from EEG end-to-end~\cite{lee2024fesde,lee2025phoneme}. These models are trained with waveform- or spectral-level objectives and evaluated using acoustic measures such as Mel Cepstral Distortion and Mel-Correlation~\cite{lee2024fesde,lee2025phoneme}. While these metrics capture signal fidelity, they offer limited insight into whether the reconstructed speech preserves semantic content. As a result, EEG-to-speech generation has been largely framed as an acoustic reconstruction problem, leaving open whether semantic information in EEG is being captured at all.

These observations suggest that the acoustic-only framing is incomplete. First, FESDE~\cite{lee2024fesde} reports that using all EEG channels outperforms restricting input to the temporal and parietal regions associated with auditory processing~\cite{sharon2019empirical}. If the task were purely acoustic, auditory-related regions should suffice; the improvement from broader channel usage suggests that EEG-to-speech generation leverages information beyond auditory processing, though the underlying mechanism remains unclear. Second, existing models overlook the data's structure. The standard benchmark, the N400 corpus~\cite{toffolo2022}, contains sentences with semantically congruent or incongruent endings designed to elicit the N400 event-related potential, a well-established index of lexical and conceptual semantic processing~\cite{kutas1980,kutas2011}. However, prior models~\cite{lee2024fesde,lee2025phoneme} treat congruent and incongruent trials identically across all training objectives, missing the opportunity to use the congruency distinction for semantic alignment. Taken together, EEG-to-speech generation involves both acoustic and semantic information, and that the paradigm's semantic structure could be used to ground EEG representations in linguistic meaning.

%This formulation leaves a key empirical observation unexplained. FESDE~\cite{lee2024fesde} reports that using all EEG channels yields better performance than restricting the input to temporal and parietal channels, even though those regions are most directly associated with auditory processing~\cite{sharon2019empirical}. Although this work suggests contributions from non-auditory regions, the underlying mechanism remains unclear. If EEG-to-speech generation were purely an acoustic task, signals from auditory-related regions should suffice. The improvement from broader channel usage suggests that EEG-to-speech generation relies on information beyond auditory processing. In addition, existing EEG-to-speech models overlook the structure of the data they are trained on. The standard benchmark, the N400 corpus~\cite{toffolo2022}, consists of sentences with semantically congruent or incongruent endings and is specifically designed to elicit the N400 event-related potential, a well-established index of lexical and conceptual semantic processing~\cite{kutas1980,kutas2011}. However, prior models~\cite{lee2024fesde,lee2025phoneme} treat congruent and incongruent trials identically, ignoring the paradigm’s semantic structure. This distinction is critical: congruent trials reflect coherent semantic processing of the target word, whereas incongruent trials disrupt this process.Leveraging this structure enables grounding EEG representations in semantic content via congruent trials, while still utilizing all trials for acoustic learning.

We address these gaps along two axes. First, for semantic grounding, we align EEG representations with a pretrained CLIP space~\cite{radford2021learning}, restricted to congruent trials so that the alignment target reflects the brain's coherent processing of the target word. CLIP is well-suited to this setting because the N400 paradigm fixes the sentence frame and varies only the final word, concentrating semantic differences at the sentence-final position; its sentence-level embedding therefore captures variations that directly correspond to the semantic content being processed by the brain. Although CLIP is not trained on neural signals, prior work has shown that its embedding space captures structure also represented in the brain~\cite{huth2016natural,pereira2018toward,tang2023semantic}. Second, for spatial structure, we model EEG as a graph defined by electrode geometry, in contrast to existing EEG-to-speech encoders that treat channels as unordered feature dimensions~\cite{krishna2020speech,lee2024fesde,lee2025phoneme,kommineni2024knowledge}. Graph-based EEG models have been explored for classification tasks such as emotion recognition~\cite{song2018eeg,zhong2020eeg}, but to our knowledge, have not been applied to the EEG-to-speech generation. Together, these design choices yield representations that jointly capture neural activity across multiple functionally distinct regions and high-level semantic content, enabling the model to leverage information beyond auditory regions, as we confirm in Section~\ref{sec:analysis}.

We propose \textbf{SENSE} (\textbf{S}emantic-\textbf{E}EG \textbf{N}eural \textbf{S}peech Synth\textbf{E}sis), a framework for EEG-to-speech generation that integrates spatial modeling with paradigm-aware semantic conditioning. SENSE includes a graph-based EEG encoder that reflects electrode geometry and an EEG Semantic Conditioning (ESC) module that aligns EEG representations with a pretrained semantic space using congruent trials, incorporating semantic information into speech generation. Figure~\ref{fig:teaser} illustrates the motivation of our work.
We summarize our contributions as follows:
\begin{itemize}
    \item We reformulate EEG-to-speech generation to incorporate both acoustic and semantic information, showing that semantic structure benefits both training and evaluation.
    \item We propose \textbf{SENSE}, a framework that combines a graph-based EEG encoder with EEG Semantic Conditioning (ESC), leveraging the congruency structure of N400-style corpora to align EEG representations with a pretrained semantic space.
    \item We show that SENSE improves both reconstruction quality and semantic consistency, achieving strong gains in unseen-subject generalization, and surpassing the strongest baseline trained on all 18 subjects in word error rate even when trained on only two subjects.

\end{itemize}
\section{Related Work}
\label{sec:related}

\noindent\textbf{Neural Speech and Language Decoding.}
Decoding speech and language from brain activity has advanced across both invasive and non-invasive modalities. Intracortical and ECoG-based approaches achieve strong performance, including continuous speech synthesis~\cite{anumanchipalli2019speech}, high-throughput neuroprostheses~\cite{moses2021neuroprosthesis,willett2023high,metzger2023high}, and low-resource ECoG-to-speech synthesis via wav2vec~2.0~\cite{baevski2020wav2vec} transfer learning~\cite{kim2023braintalker}, but all require invasive recordings~\cite{ullsperger2010simultaneous}. Non-invasive alternatives are safer but operate on weaker signals: fMRI enables language reconstruction with language-model priors~\cite{tang2023semantic}, while MEG and EEG offer high temporal resolution~\cite{defossez2023decoding}. Several recent non-invasive works target \emph{text} rather than waveforms (via contrastive retrieval against wav2vec~2.0~\cite{defossez2023decoding}, discrete EEG codex with pretrained BART~\cite{duan2023dewave}, or Whisper-based MEG-to-text decoding~\cite{yang2026neuspeech}) and thus differ from our task in either modality or output. Within EEG-to-speech \emph{waveform generation}, early work predicted acoustic features~\cite{krishna2020speech}, while recent end-to-end models reconstruct waveforms directly~\cite{lee2024fesde,lee2025phoneme}; these are the comparable systems on the N400 corpus~\cite{toffolo2022} and serve as our baselines. All formulate the task as acoustic reconstruction, despite evidence that using all channels improves performance beyond auditory regions~\cite{lee2024fesde,sharon2019empirical}. In contrast, we treat EEG-to-speech generation as involving both acoustic and semantic information, and incorporate semantic structure into both training and evaluation.

%Decoding speech from brain activity has advanced across both invasive and non-invasive modalities. Intracortical and ECoG-based approaches have achieved strong performance, including continuous speech synthesis~\cite{anumanchipalli2019speech} and high-throughput neuroprostheses~\cite{moses2021neuroprosthesis,willett2023high,metzger2023high}, but require invasive recordings~\cite{ullsperger2010simultaneous}. Non-invasive approaches provide safer alternatives: fMRI enables language reconstruction with language-model priors~\cite{tang2023semantic}, while MEG and EEG offer high temporal resolution for speech decoding~\cite{defossez2023decoding}. In EEG-to-speech generation, early work predicted acoustic features~\cite{krishna2020speech}, while recent models reconstruct waveforms end-to-end~\cite{lee2024fesde,lee2025phoneme}. These approaches formulate the task as acoustic reconstruction, despite evidence that using all channels improves performance beyond auditory regions~\cite{lee2024fesde,sharon2019empirical}.

\noindent\textbf{Semantic Representations from Brain Activity.}
Semantic representations capture high-level meaning beyond sensory details. Prior work shows that such information is encoded in brain activity: semantic organization across cortex has been mapped during natural speech~\cite{huth2016natural}, and conceptual meaning can be decoded from fMRI~\cite{pereira2018toward}. The N400 component further indicates that semantic processing is accessible at the EEG level~\cite{kutas1980,kutas2011}. More recently, brain signals have been aligned with pretrained embedding spaces: fMRI and EEG have been mapped into the CLIP~\cite{radford2021learning} for visual retrieval and reconstruction~\cite{takagi2023high,scotti2023reconstructing,scotti2024mindeye2,song2023decoding,li2024visual}, and MEG/EEG have been aligned with wav2vec~2.0~\cite{baevski2020wav2vec} to identify corresponding speech segments from brain activity~\cite{defossez2023decoding}. These results suggest that pretrained embedding spaces, despite not being trained on neural data, can serve as effective targets for extracting semantically meaningful information from brain signals via alignment. Unlike alignment-based approaches focused on retrieval or identification, we use semantic alignment as a conditioning mechanism within a generative pipeline that directly synthesizes speech waveforms from EEG.

\noindent\textbf{EEG Representation Learning.}
EEG representation learning is challenging due to low signal-to-noise ratio and inter-subject variability, and limited data~\cite{lotte2018review,roy2019deep,chen2022toward}. Convolutional models such as EEGNet~\cite{lawhern2018eegnet} treat channels as feature dimensions processed by temporal and spatial filters and existing EEG-to-speech models follow this convention without explicitly modeling spatial relationships~\cite{krishna2020speech,lee2024fesde,lee2025phoneme,kommineni2024knowledge}. Graph-based approaches model electrode geometry with learned or biologically motivated adjacencies~\cite{song2018eeg,zhong2020eeg}, but are primarily applied to classification tasks such as emotion recognition. Large-scale self-supervised pretraining has also been explored to improve generalization~\cite{kostas2021bendr,jiang2024large}. In contrast, we apply graph-based spatial modeling to a generative setting rather than classification, and incorporate structure not through within-modality pretraining but via alignment with a semantic space learned from large-scale multimodal data.
\section{Method}
\label{sec:method}

\begin{figure}
    \centering
    \includegraphics[width=0.9\textwidth]{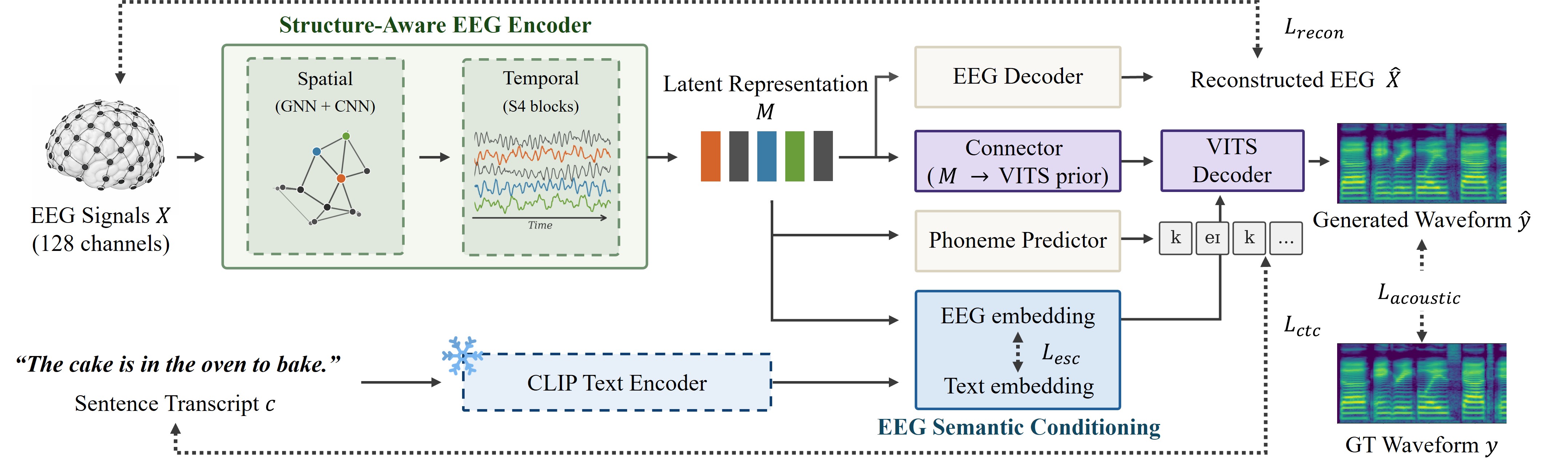}
    \vspace{-3mm}
    \caption{\textbf{SENSE training pipeline.} EEG signal $\mathbf{X}$ is encoded by a structure-aware EEG encoder into a latent $\mathbf{M}$. $\mathbf{M}$ is mapped to the VITS prior via a connector module and conditioned at the decoder using an EEG embedding. The EEG embedding is aligned with a frozen CLIP text embedding (ESC, congruent trials only). Solid arrows indicate the forward data flow, while dashed arrows represent loss supervision, connecting model predictions to their corresponding training targets.}
    \label{fig:overview}
        \vspace{-7mm}
\end{figure}

Figure~\ref{fig:overview} illustrates the overall pipeline. 
Each training example consists of an $\mathbf{X} \in \mathbb{R}^{128 \times T}$, the corresponding sentence transcript $c$, and the recorded speech audio $y$. EEG is encoded by the EEG Module (Section~\ref{sec:eeg_module}) into a latent representation $\mathbf{M} \in \mathbb{R}^{d \times T'}$ ($d{=}192$), which is then mapped to the prior of a VITS speech decoder~\citep{kim2021conditional}, comprising a normalizing flow and a HiFi-GAN vocoder~\citep{kong2020hifi}. The decoder is supervised using $y$ through the standard VITS training objectives. During training, $\mathbf{M}$ is further guided by two supervisory signals derived from $c$: (1) phoneme sequence supervision, where phoneme targets are obtained by tokenizing $c$ (Section~\ref{sec:ctc}), and (2) EEG Semantic Conditioning, where $c$ is used as input to a frozen CLIP model (Section~\ref{sec:esc}). At inference time, only EEG is required. A semantic conditioning vector $\mathbf{g}_{\mathrm{eeg}} \in \mathbb{R}^{d}$, derived from $\mathbf{M}$, is used to directly condition the decoder.

SENSE addresses two limitations identified in Section~\ref{sec:intro}. First, prior EEG-to-speech models treat EEG channels as unordered feature dimensions, ignoring their spatial structure; we address this with a graph-based encoder defined over electrode geometry. Second, prior models use the N400 corpus in a paradigm-agnostic manner, treating congruent and incongruent trials identically; we address this with EEG Semantic Conditioning (ESC), which restricts semantic alignment to congruent trials. 
%------------------------------------------------------------------

\subsection{Structure-Aware EEG Encoder}
\label{sec:eeg_module}
Prior EEG-to-speech encoders are typically trained solely with a self-reconstruction or phoneme objectives~\citep{kommineni2024knowledge,lee2024fesde,lee2025phoneme}, producing representations optimized primarily for acoustic fidelity.
In contrast, SENSE jointly supervises the EEG module with both phonemic and semantic objectives, yielding a latent representation that captures acoustic structure, phonetic content, and high-level semantic meaning. The module consists of a spatial encoder that exploits electrode geometry, a temporal encoder that models long-range dependencies, and two auxiliary components used only during training.

\noindent\textbf{Spatial Encoder.}
EEG electrodes are arranged according to standard scalp positions, and signals originating from spatially proximate cortical sources exhibit strong correlations~\citep{nunez2006electric}. Conventional 1D CNN encoders overlook this spatial structure by treating channels independently. To address this limitation, we employ a graph-based spatial encoder that explicitly models electrode topology. The encoder integrates three components: a learnable channel gate, a graph convolution over electrode geometry, and a parallel CNN skip connection.

{\tiny\textbullet}  \textit{Channel attention gate:} A learnable per-channel sigmoid gate, $\mathbf{X}_{\mathrm{att}} = \mathbf{X} \odot \sigma(\mathbf{w})$, where 
$\mathbf{w} \in \mathbb{R}^{128}$ is a learnable parameter and $\sigma$ denotes 
the sigmoid function, reweights each electrode prior to spatial mixing, with a positive bias on auditory-related electrodes~\citep{sharon2019empirical} as a warm-start.

{\tiny\textbullet} \textit{Graph convolution:} The 128 electrodes are represented as graph nodes with 3D coordinates defined by the BioSemi montage~\citep{gramfort2013meg}. We construct a fixed adjacency $\mathbf{A} \in \mathbb{R}^{128 \times 128}$ using a Gaussian kernel over pairwise Euclidean distances. 
At each time step, node features are updated through two layers of residual graph convolution: $\mathbf{h}_{i}^{(\ell)} = \mathbf{h}_{i}^{(\ell-1)} + f^{(\ell)}\!\left(\sum_{j} A_{ij}\,\mathbf{h}_{j}^{(\ell-1)}\right)$, $\ell \in \{1, 2\}$, where each $f^{(\ell)}$ is a learnable feedforward block. The resulting node representations are then averaged over all 128 nodes and projected to dimension $d$, yielding $\mathbf{H}_{\mathrm{GNN}} \in \mathbb{R}^{d \times T}$. We adopt a fixed anatomy-derived adjacency rather than learning the graph structure from data~\citep{velivckovic2017graph,song2018eeg}; with only 7,200 training pairs, joint optimization with the downstream task is prone to overfitting.

{\tiny\textbullet} \textit{CNN skip connection:} Since the graph convolution involves aggregation over all 128 nodes, it can attenuate channel-specific information. To mitigate this, we introduce a parallel CNN branch applied to $\mathbf{X}_{\mathrm{att}}$, which captures local channel interactions through learned convolutions, producing $\mathbf{H}_{\mathrm{CNN}} \in \mathbb{R}^{d \times T}$. The outputs of two branches are combined via a learnable scalar gate $g \in \mathbb{R}$: $\mathbf{H} = \mathbf{H}_{\mathrm{GNN}} + \sigma(g) \cdot \tilde{\mathbf{H}}_{\mathrm{CNN}}$, where $\tilde{\mathbf{H}}_{\mathrm{CNN}}$ denotes $\mathbf{H}_{\mathrm{CNN}}$ rescaled to match the variance of $\mathbf{H}_{\mathrm{GNN}}$. The gate $g$ is initialized to favor the graph branch during the early training.

\noindent\textbf{Temporal Encoder.}
$\mathbf{H}$ is temporally downsampled and processed by four S4 blocks~\citep{gu2021efficiently}, yielding $\mathbf{M} \in \mathbb{R}^{d \times T'}$. We use S4 because EEG sequences remain long after downsampling; FESDE~\citep{lee2024fesde} adopts the same backbone for fair comparison.

\noindent\textbf{Auxiliary Components.}
Following FESDE~\citep{lee2024fesde}, we attach two auxiliary components used only during training. An \textit{EEG decoder} (lightweight transposed convolutions) reconstructs the input signal from $\mathbf{M}$ under a length-masked cosine loss $\mathcal{L}_{\mathrm{recon}}$ in~\eqref{eq:total_loss}. A \textit{connector}, consisting of a self-attention encoder followed by a linear projection, maps $\mathbf{M}$ into the mean and log-variance of the VITS prior. 
A stop-gradient between the EEG module and the connector prevents VITS prior-matching gradients from dominating the EEG encoder, so $\mathbf{M}$ is shaped jointly by reconstruction, phoneme, and semantic objectives. The EEG decoder is discarded at inference; the connector remains. Implementation details are in Appendix~\ref{app:impl}.

%------------------------------------------------------------------
\subsection{Phoneme Sequence Supervision}
\label{sec:ctc}

Mel-spectrogram reconstruction provides holistic acoustic supervision but does not explicitly encourage $\mathbf{M}$ to encode categorical phonemic structure. Prior work shows that auxiliary phoneme prediction significantly improves EEG-to-speech performance~\cite{lee2025phoneme}. We adopt this approach with a Conformer block~\cite{gulati2020conformer} followed by an LSTM attention decoder over a 51-symbol IPA vocabulary, trained with a CTC loss~\cite{graves2006connectionist} against the phoneme sequence tokenized from $c$. The predictor is discarded at inference. Implementation details are in Appendix~\ref{app:ctc}.

%------------------------------------------------------------------
\subsection{EEG Semantic Conditioning}
\label{sec:esc}

Phoneme supervision encourages $\mathbf{M}$ to encode fine-grained phonetic structure, but leaves a mismatch: the decoder is conditioned on text during training, yet no such signal is available at inference. We resolve this with EEG Semantic Conditioning (ESC), which learns to map the EEG latent into the decoder’s conditioning space and replaces text-based conditioning at inference. Unlike prior brain-to-CLIP alignment work~\citep{takagi2023high,scotti2023reconstructing,song2023decoding} targeting retrieval or reconstruction, ESC is designed for generative conditioning via two choices: a \textit{congruency restriction} and a \textit{conditioning bridge} that exposes $\mathbf{g}_{\mathrm{eeg}}$ to the decoder during training so it remains in-distribution at inference.

\noindent\textbf{Conditioning-Space Alignment.}
The speech decoder conditions its flow and HiFi-GAN vocoder via a learned 
projection $W_{\mathrm{clip}}\colon \mathbf{c}_{\mathrm{text}} \in \mathbb{R}^{512} 
\to \mathbf{g}_{\mathrm{clip}} \in \mathbb{R}^{d}$ of CLIP text embeddings, 
where $\mathbf{c}_{\mathrm{text}}$ is the output of a frozen CLIP ViT-B/32 
encoder applied to the sentence $c$. To enable text-free conditioning, we learn an additional linear map $W_{\mathrm{eeg}} \in \mathbb{R}^{d \times d}$ that projects the temporal mean of $\mathbf{M}$ into the same conditioning space, yielding $\mathbf{g}_{\mathrm{eeg}} \in \mathbb{R}^{d}$. The alignment target is $\mathbf{g}_{\mathrm{clip}} = \mathrm{sg}[W_{\mathrm{clip}}(\mathbf{c}_{\mathrm{text}})]$, where $\mathrm{sg}[\cdot]$ denotes stop-gradient; this prevents $\mathcal{L}_{\mathrm{esc}}$ from pulling $W_{\mathrm{clip}}$ toward the EEG-reachable subspace and degrading conditioning quality, so $W_{\mathrm{eeg}}$ targets a fixed space shaped solely by acoustic objectives.

\noindent\textbf{Congruency-Aware Cosine Loss.}
The N400 corpus contains congruent and incongruent sentence-final words. Aligning the EEG representation to semantic content is meaningful only for congruent trials, where neural responses reflect coherent semantic processing of the target word~\citep{kutas1980,kutas2011}. Let $\mathcal{C}$ denote the subset of congruent indices in the batch. The ESC loss minimizes cosine distance in the conditioning space restricted to $\mathcal{C}$:
%\begin{equation}
$\mathcal{L}_{\mathrm{esc}} =
    \frac{1}{|\mathcal{C}|}\sum_{i \in \mathcal{C}}
    \left(1 - \frac{\mathbf{g}_{\mathrm{eeg},i}^{\top}\,\mathbf{g}_{\mathrm{clip},i}}
                   {\|\mathbf{g}_{\mathrm{eeg},i}\|\,\|\mathbf{g}_{\mathrm{clip},i}\|}
    \right)$.
%\label{eq:esc}
%\end{equation}
All other training objectives in Section~\ref{sec:loss} use both congruent and incongruent trials, so incongruent trials contribute to acoustic and phonemic learning without injecting noise into semantic alignment.

\noindent\textbf{Conditioning Bridge.}
To ensure that the decoder accepts $\mathbf{g}_{\mathrm{eeg}}$ at inference, we expose it during training. The decoder is conditioned on three sources, each sampled with roughly equal probability: $\mathbf{g}_{\mathrm{eeg}}$, $W_{\mathrm{clip}}(\mathbf{c}_{\mathrm{text}})$, and a learnable null embedding (analogous to unconditional dropout in classifier-free guidance training~\citep{ho2022classifier}, though no guided sampling is performed at inference). When conditioning on $\mathbf{g}_{\mathrm{eeg}}$, gradients are blocked at both $\mathbf{M}$ and $W_{\mathrm{eeg}}$, so acoustic gradients do not leak back into the EEG encoder and $W_{\mathrm{eeg}}$ is optimized solely by $\mathcal{L}_{\mathrm{esc}}$.
This ensures $\mathbf{g}_{\mathrm{eeg}}$ is in-distribution at inference and the decoder remains compatible with text-based conditioning during training.

%------------------------------------------------------------------
\subsection{Training Objective and Inference}
\label{sec:loss}

\noindent\textbf{Three Complementary Supervisory Signals.}
SENSE supervises $\mathbf{M}$ at three levels of abstraction. EEG self-reconstruction ($\mathcal{L}_{\mathrm{recon}}$) preserves low-level signal fidelity, phoneme supervision ($\mathcal{L}_{\mathrm{ctc}}$) imposes discrete phonetic structure, and semantic conditioning ($\mathcal{L}_{\mathrm{esc}}$) injects high-level meaning by aligning the latent with a pretrained semantic space. These signals operate at distinct granularities (raw signal, phonemes, sentence-level meaning) and are complementary: $\mathbf{M}$ must simultaneously support EEG reconstruction, phoneme prediction, and semantic alignment. 

\noindent\textbf{Total Loss.} We define the total loss as:
\begin{equation}
  \mathcal{L} =
    \underbrace{%
      \lambda_{\mathrm{mel}}\mathcal{L}_{\mathrm{mel}}
      + \mathcal{L}_{\mathrm{adv}}
      + \mathcal{L}_{\mathrm{fm}}
      + \lambda_{\mathrm{kl}}\mathcal{L}_{\mathrm{kl}}
    }_{\mathcal{L}_{\text{acoustic (VITS)}}}
    + \underbrace{\lambda_{\mathrm{recon}}\mathcal{L}_{\mathrm{recon}}}_{\text{EEG self-recon}}
    + \underbrace{\lambda_{\mathrm{ctc}}\mathcal{L}_{\mathrm{ctc}}}_{\text{phoneme}}
    + \underbrace{\lambda_{\mathrm{esc}}\mathcal{L}_{\mathrm{esc}}}_{\text{semantic}},
  \label{eq:total_loss}
\end{equation}
with $\{\lambda_{\mathrm{mel}}, \lambda_{\mathrm{kl}}, \lambda_{\mathrm{recon}}, \lambda_{\mathrm{ctc}}, \lambda_{\mathrm{esc}}\}{=}\{45,1.0,1.0,0.3,0.5\}$. $\mathcal{L}_{\mathrm{mel}}$ (mel-spectrogram L1 reconstruction), $\mathcal{L}_{\mathrm{adv}}$, $\mathcal{L}_{\mathrm{fm}}$, and $\mathcal{L}_{\mathrm{kl}}$ follow VITS~\cite{kim2021conditional}. $\mathcal{L}_{\mathrm{recon}}$ is the cosine reconstruction loss applied to the EEG decoder output (Section~\ref{sec:eeg_module}), $\mathcal{L}_{\mathrm{ctc}}$ is the phoneme prediction loss of FE-Phoneme~\cite{lee2025phoneme} (Section~\ref{sec:ctc}), and $\mathcal{L}_{\mathrm{esc}}$ is the congruency-restricted alignment loss defined in Section~\ref{sec:esc}.

\noindent\textbf{Inference.}
The EEG decoder, phoneme predictor, and CLIP encoder are discarded. Raw EEG is processed by the EEG encoder to produce $\mathbf{M}$, which is mapped to $\mathbf{g}_{\mathrm{eeg}}$ via $W_{\mathrm{eeg}}$ and to the VITS prior via the Connector. The decoder synthesizes the waveform conditioned on $\mathbf{g}_{\mathrm{eeg}}$, without text or guided sampling; the null embedding from training is unused at inference and serves only as regularization.

\section{Experiments}
\label{sec:experiments}
\vspace{-3mm}

\noindent\textbf{Dataset.} We evaluated on the N400 EEG corpus~\cite{toffolo2022}, which comprises 24 subjects, each with 440 sentences, recorded using 128-channel EEG at 512\,Hz. Following \cite{lee2024fesde,lee2025phoneme}, we excluded four unreliable subjects, leaving 20 subjects total, and used a standard preprocessing pipeline; speech was downsampled to 22{,}050\,Hz with 80-band mel-spectrogram targets (Appendix~\ref{app:preprocessing}). The training set contains 7{,}200 EEG-speech pairs from 18 subjects $\times$ 400 sentences. We evaluated performance under three test conditions following~\cite{lee2024fesde}: \textit{Unseen audio} (training subjects, held-out sentences; 720 pairs), \textit{Unseen subject} (held-out subjects sub-23, sub-24 with training sentences; 800 pairs), and \textit{Unseen both} (held-out subjects with held-out sentences; 80 pairs). For cross-subject scaling experiments, we used a nested subset structure $k{=}2 \subset k{=}4 \subset k{=}8 \subset k{=}18$ (Appendix~\ref{app:training}).

\noindent\textbf{Baselines and Metrics.} 
We compared SENSE against FESDE~\cite{lee2024fesde} and FE-Phoneme~\cite{lee2025phoneme}, the comparable existing end-to-end EEG-to-speech models on this corpus, retrained using their official implementations. Acoustic fidelity was measured by MCD, Mel-Correlation, and STOI; linguistic content by WER (via Whisper-large~\cite{radford2023robust}); rescaled BERTScore (BERT)~\cite{zhang2019bertscore}, and CLIP-Sim~\cite{radford2021learning}. WER may exceed 1 due to insertion errors. Since incongruent trials are designed to violate the sentence-final semantic expectation, semantic metrics were computed only on congruent trials; acoustic metrics used all trials.

\noindent\textbf{Training Setup.} SENSE was trained for 200k iterations until validation performance plateaus, with loss weights $\lambda_{\mathrm{ctc}}=0.3$ and $\lambda_{\mathrm{esc}}=0.5$ selected via the sensitivity analysis in Appendix~\ref{app:lambda_sensitivity}. Baselines were trained for 100k iterations as specified in their official implementations and original papers, with extended training providing no further gains. All results were averaged over five seeds. Implementation details are in Appendix~\ref{app:impl}.

% ===main table
\begin{table*}[t]
\centering
\caption{\textbf{Comparison of end-to-end EEG-to-speech models} across three evaluation conditions. }
\label{tab:main_results}
\setlength{\tabcolsep}{8pt}
\renewcommand{\arraystretch}{0.7}
\scriptsize{
\begin{tabular}{llcccccc}
\toprule
& \textbf{Model}
  & \textbf{MCD}$\downarrow$
  & \textbf{Mel-Corr}$\uparrow$
  & \textbf{STOI}$\uparrow$
  & \textbf{WER}$\downarrow$
  & \textbf{BERT}$\uparrow$
  & \textbf{CLIP-Sim}$\uparrow$\\
\midrule
\multirow{3}{*}{\rotatebox{90}{\scriptsize Both}}
& FESDE~\cite{lee2024fesde}
  & 11.80\,{\tiny$\pm$0.12} & 19.87\,{\tiny$\pm$1.97}  & 0.2785 & 1.2186 & 0.0381 & 0.7557\\
& FE-Phoneme~\cite{lee2025phoneme}
  & 10.52\,{\tiny$\pm$0.08} & 28.29\,{\tiny$\pm$1.34} & 0.3979 & 1.1610 & 0.0428 & 0.7509 \\
& \textbf{SENSE}
  & \textbf{10.44}\,{\tiny$\pm$0.10} & \textbf{31.18}\,{\tiny$\pm$0.60}  & \textbf{0.3996} & \textbf{1.0186} & \textbf{0.0852} & \textbf{0.7654}\\
\midrule
\multirow{3}{*}{\rotatebox{90}{\scriptsize Audio}}
& FESDE~\cite{lee2024fesde}
  & 11.71\,{\tiny$\pm$0.04} & 19.14\,{\tiny$\pm$0.47}  & 0.2792 & 1.2125 & 0.0289 & \textbf{0.7601}  \\
& FE-Phoneme~\cite{lee2025phoneme}
  & 10.73\,{\tiny$\pm$0.06} & 26.53\,{\tiny$\pm$0.30}  & 0.3918 & 1.1360 & 0.0602 & 0.7554\\
& \textbf{SENSE}
  & \textbf{10.44}\,{\tiny$\pm$0.01} & \textbf{28.50}\,{\tiny$\pm$0.24} & \textbf{0.4039} & \textbf{1.0375} & \textbf{0.0959} & \textbf{0.7601}\\
\midrule
\multirow{3}{*}{\rotatebox{90}{\scriptsize Subject}}
& FESDE~\cite{lee2024fesde}
  & 11.73\,{\tiny$\pm$0.04} & 18.70\,{\tiny$\pm$0.29} & 0.2744 & 1.2171& 0.0269 & 0.7587 \\
& FE-Phoneme~\cite{lee2025phoneme}
  & 10.65\,{\tiny$\pm$0.02} & 26.90\,{\tiny$\pm$0.08} & 0.3806 & 1.1231 & 0.0564 & 0.7555\\
& \textbf{SENSE}
  & \textbf{9.928}\,{\tiny$\pm$0.03} & \textbf{32.07}\,{\tiny$\pm$0.14} & \textbf{0.4126} & \textbf{0.9948} & \textbf{0.1088} & \textbf{0.7647} \\
\bottomrule
\end{tabular}
}
    \vspace{-3mm}
\end{table*}

% ===== melspectrogram
\begin{figure}[h!]
    \centering
    \includegraphics[width=0.9\linewidth]{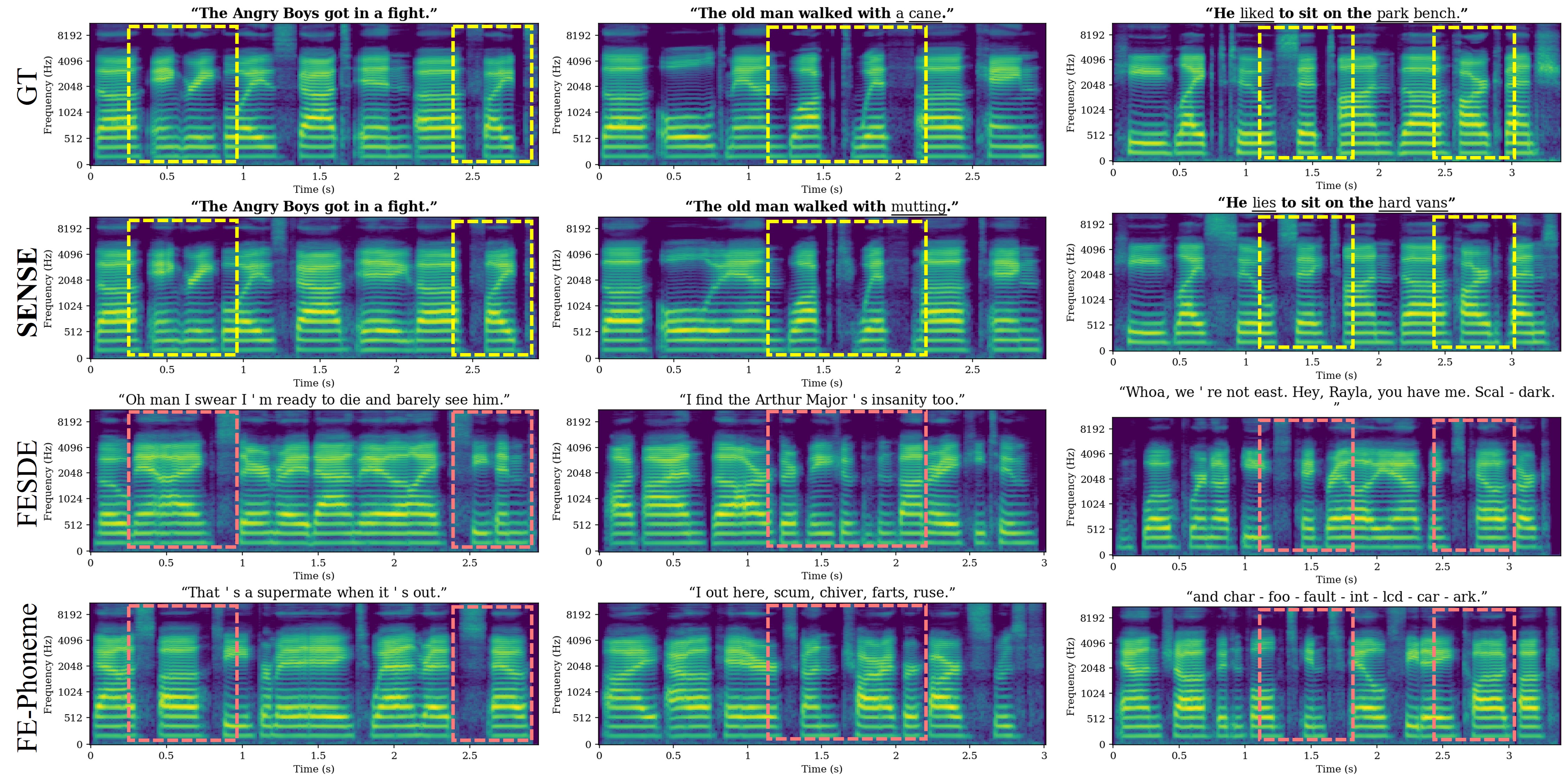}
    \vspace{-4mm}
    \caption{\textbf{Qualitative comparison of generated speech.} Mel-spectrograms and Whisper-large transcriptions for ground truth, SENSE, FESDE, and FE-Phoneme. Words matching the ground truth are shown in \textbf{bold}, while semantically or phonetically similar words are \underline{underlined}.}
    \label{fig:spectrogram}
        \vspace{-5mm}
\end{figure}

% %=== Top 20 %
% \begin{table}[t!]
% \centering
% \caption{\textbf{Performance on the 80 congruent test samples with the lowest per-sample WER under SENSE.} All models are evaluated on the same sample indices. As this subset is defined by SENSE’s own performance, it does not represent a fair head-to-head comparison; instead, it serves to assess whether the gains arise from broadly informative EEG signals or SENSE-specific recovery.}
% \label{tab:semantic_top20}
%     \vspace{-2mm}
% \setlength{\tabcolsep}{30pt}
% \renewcommand{\arraystretch}{0.7}
% \scriptsize{
% \begin{tabular}{lccc}
% \toprule
% \textbf{Model} & \textbf{WER} $\downarrow$ & \textbf{BERTScore} $\uparrow$ & \textbf{CLIP-Sim} $\uparrow$ \\
% \midrule
% FESDE~\cite{lee2024fesde}          & 1.136 & 0.840 & 0.765 \\
% FE-Phoneme~\cite{lee2025phoneme} & 1.061 & 0.842 & 0.763 \\
% \midrule
% SENSE                & \textbf{0.773} & \textbf{0.860} & \textbf{0.790} \\
% \bottomrule
% \end{tabular}
% }
%     \vspace{-3mm}
% \end{table}

% ===== Scaling curve
\begin{figure}[t]
    \centering
    \includegraphics[width=0.9\linewidth]{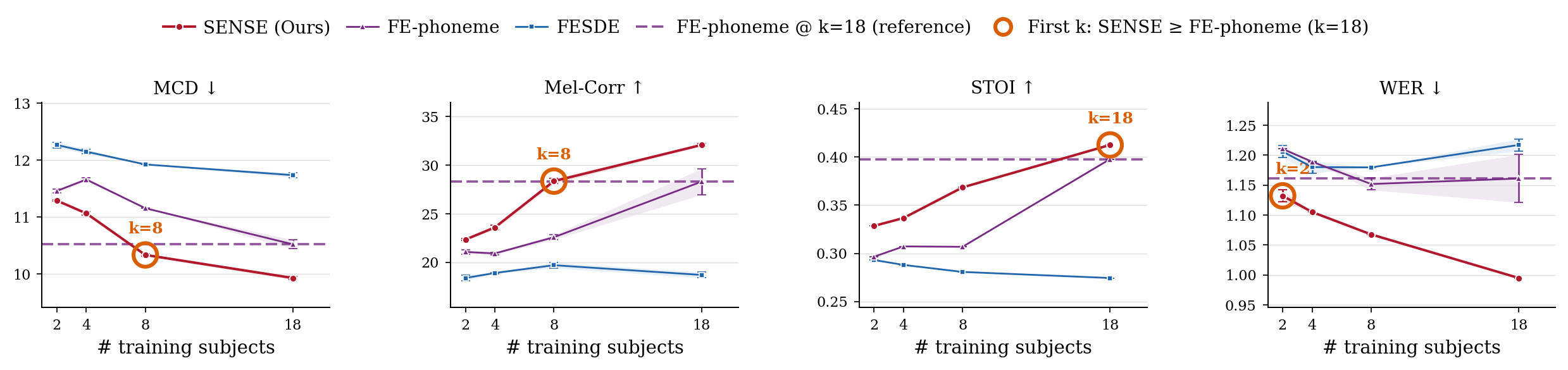}
        \vspace{-4mm}
        \caption{\textbf{Scaling with the number of training subjects.} Models are trained on $k \in \{2, 4, 8, 18\}$ subjects and evaluated on the held-out pair (sub-23, sub-24). Shaded regions: $\pm 1$ std over 5 seeds. Dashed line: FE-Phoneme at $k{=}18$ (best baseline). Orange circles: the smallest $k$ at which SENSE first matches or exceeds this reference.}
    \label{fig:scaling_curve}
            \vspace{1mm}
%\end{figure}
%\begin{figure}[t]
    \centering
    \includegraphics[width=0.9\linewidth]{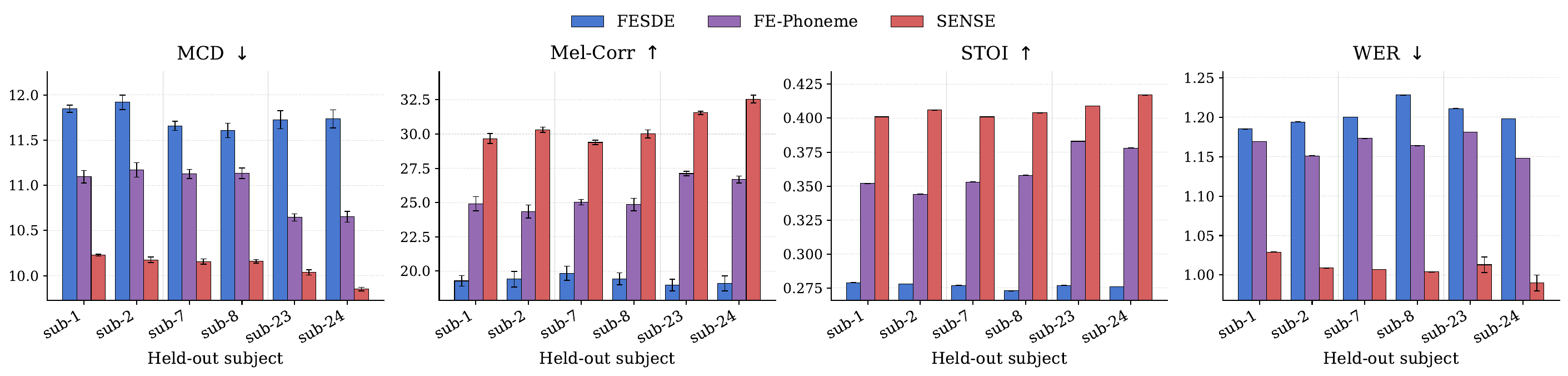}
    \vspace{-4mm}
    \caption{\textbf{Per-subject results across various held-out pairs.} All models are retrained from scratch on the remaining 18 subjects for each pair. Error bars: $\pm 1$ std over 5 seeds.}
    \label{fig:cross_subject}
                \vspace{-5mm}
\end{figure}

\vspace{-3mm}
\subsection{Main Results}
\label{sec:main}
\vspace{-2mm}
SENSE outperforms both baselines across all splits and metrics (Table~\ref{tab:main_results}), with the largest gains on the \textit{unseen subject} split, indicating that it learns representations that generalize across subjects rather than relying on subject-specific cues. Absolute WER values near 1 nonetheless show that intelligibility-level reconstruction from EEG remains an open challenge. BERTScore and CLIP-Sim differences are smaller, as these sentence-level metrics are less sensitive to fine-grained lexical differences than WER. Qualitative examples in Figure~\ref{fig:spectrogram} show that SENSE produces speech that is both acoustically consistent and semantically meaningful, while baselines often fail to recover lexical content. \emph{For additional qualitative evidence, please refer to the \textbf{audio samples} in the supplementary video and Appendix~\ref{app:additional_samples}.}

% ==== Ablation

\begin{table}[t]
\centering
\caption{\textbf{Ablation study of the proposed SENSE components}. Each row corresponds to removing one component from the full model. w shuffled ESC uses the same ESC loss and conditioning bridge as SENSE, but randomly permutes CLIP targets within each batch.}
\label{tab:ablation}
                \vspace{-2mm}
\setlength{\tabcolsep}{4pt}
\renewcommand{\arraystretch}{0.7}
\scriptsize{
\begin{tabular}{l|ccccc|cccccc}
\toprule
\textbf{Model}
  & \textbf{Ch.Att} & \textbf{GNN} & \textbf{CNN Skip} & \textbf{CTC} & \textbf{ESC}
  & \textbf{MCD}$\downarrow$
  & \textbf{Mel-Corr}$\uparrow$
  & \textbf{STOI}$\uparrow$
  & \textbf{WER}$\downarrow$
  & \textbf{BERT}$\uparrow$
  & \textbf{CLIP-Sim}$\uparrow$ \\
\midrule
\textbf{SENSE}
  & \checkmark & \checkmark & \checkmark & \checkmark & \checkmark
  & \textbf{9.93}\,{\tiny$\pm$0.03}  & \textbf{32.07}\,{\tiny$\pm$0.14} & \textbf{0.4126} & \textbf{0.9948} & \textbf{0.1088} & \textbf{0.7647} \\
\midrule
w/o ESC
& \checkmark & \checkmark & \checkmark & \checkmark & -
  & 10.34\,{\tiny$\pm$0.03} & 29.43\,{\tiny$\pm$0.27} & 0.4016 & 1.0323 & 0.0763 & 0.7601 \\
w shuffled ESC
& \checkmark & \checkmark & \checkmark & \checkmark & \checkmark(shuf)
 & 10.38\,{\tiny$\pm$0.04} & 29.25\,{\tiny$\pm$0.28} & 0.4002 & 1.0385 & 0.0721 & 0.7595  \\
w/o CTC
  & \checkmark & \checkmark & \checkmark & - & \checkmark
  & 10.50\,{\tiny$\pm$0.01} & 27.24\,{\tiny$\pm$0.27} & 0.3721 & 1.0488 & 0.0588 & 0.7624 \\
w/o CNN Skip
  & \checkmark & \checkmark & - & \checkmark & \checkmark
  & 10.40\,{\tiny$\pm$0.01} & 29.64\,{\tiny$\pm$0.25} & 0.3970 & 1.0306 & 0.0724 & 0.7579 \\
w/o GNN
  & \checkmark & - & - & \checkmark & \checkmark
  & 10.11\,{\tiny$\pm$0.02} & 29.54\,{\tiny$\pm$0.22} & 0.3951 & 1.0355 & 0.0788 & 0.7577 \\
w/o Ch.Att
  & - & \checkmark & \checkmark & \checkmark & \checkmark
  & 10.24\,{\tiny$\pm$0.01} & 29.08\,{\tiny$\pm$0.16} & 0.4046 & 1.0170 & 0.0926 & 0.7639 \\
\bottomrule
\end{tabular}%
}
\vspace{1mm}
%\end{table}
% ======= ESC ablation
%\begin{table}[t!]
\centering
\caption{\textbf{Effect of adding ESC to baseline encoders.} We incorporate ESC into FESDE and FE-Phoneme without other changes. While both baselines improve, they remain consistently below SENSE across all metrics, indicating that ESC complements rather than replaces the SENSE encoder.}
\label{tab:esc_ablation}
                \vspace{-2mm}
\setlength{\tabcolsep}{10pt}
\renewcommand{\arraystretch}{0.7}
\scriptsize{
\begin{tabular}{cccccccc}
\toprule
 Backbone & ESC
      & MCD $\downarrow$ & Mel-Corr $\uparrow$
      & STOI $\uparrow$ & WER $\downarrow$ & BERT $\uparrow$ & CLIP-sim $\uparrow$\\
\midrule
FESDE~\cite{lee2024fesde}  & \checkmark
  & 10.93\,{\tiny$\pm$0.03}  & 26.27\,{\tiny$\pm$0.19}  & 0.363 & 1.084 & 0.045 & 0.7500\\
FE-Phoneme~\cite{lee2025phoneme}& \checkmark
  & 10.04\,{\tiny$\pm$0.02}  & 31.17\,{\tiny$\pm$0.21}  & 0.411 & 1.102 & 0.070 & 0.7544\\
\midrule
 SENSE   & \checkmark
  & \textbf{9.93}\,{\tiny$\pm$0.03}  & \textbf{32.07}\,{\tiny$\pm$0.14}  & \textbf{0.413}
  & \textbf{0.995} & \textbf{0.109} & \textbf{0.7647}\\
\bottomrule
\end{tabular}
}
\vspace{1mm}
%\end{table}
%\begin{table}[t]
\centering
\caption{\textbf{Performance across different graph adjacency structures.} \textit{BioSemi topo}: anatomy-based adjacency derived from electrode geometry (used in SENSE). \textit{Fully connected}: uniform adjacency. \textit{Random}: fixed random adjacency. \textit{GAT}: learned edge attention via Graph Attention Network.}
\label{tab:graph_ablation}
                \vspace{-2mm}
\setlength{\tabcolsep}{11pt}
\renewcommand{\arraystretch}{0.7}
\scriptsize{
\begin{tabular}{l|lc|cccc}
\toprule
\textbf{Model}
  & \textbf{Graph Topology} & \textbf{CNN Skip}
  & \textbf{MCD}$\downarrow$
  & \textbf{Mel-Corr}$\uparrow$
  & \textbf{STOI}$\uparrow$
  & \textbf{WER}$\downarrow$ \\
\midrule
\textbf{SENSE}
  & Biosemi topo & \checkmark
  & \textbf{9.93}\,{\tiny$\pm$0.03}   & \textbf{32.07}\,{\tiny$\pm$0.14} & \textbf{0.4126}  & \textbf{0.9948}  \\
\midrule
GNN (topo, no skip)
  & Biosemi topo & --
  & 10.40\,{\tiny$\pm$0.02}  & \underline{29.64}\,{\tiny$\pm$0.25} & \underline{0.3970} & 1.0306 \\
GNN (GAT attn. topo)
  & Biosemi topo & --
  & \underline{10.21}\,{\tiny$\pm$0.03}  & 29.10\,{\tiny$\pm$0.10} & 0.3835 & \underline{1.0183} \\
GNN (fully-conn.)
  & Fully-conn.  & --
  & 10.96\,{\tiny$\pm$0.05}  & 26.17\,{\tiny$\pm$0.27} & 0.3684 & 1.0945 \\
GNN (random)
  & Random       & --
  & 10.67\,{\tiny$\pm$0.03}  & 27.15\,{\tiny$\pm$0.18} & 0.3803 & 1.0806 \\
\bottomrule
\end{tabular}%
}
\vspace{1mm}
%\end{table}
% ==== CLIP CLAP BERT table
%\begin{table}[t]
\centering
\caption{\textbf{Performance across different semantic encoders} in the ESC module.}
\label{tab:esc_encoder}
                \vspace{-2mm}
\setlength{\tabcolsep}{17pt}
\renewcommand{\arraystretch}{0.7}
\scriptsize{
\begin{tabular}{lccccc}
\toprule
Semantic Encoder & MCD $\downarrow$  & Mel-Corr $\uparrow$ & STOI $\uparrow$ & WER $\downarrow$ & BERT $\uparrow$ \\
\midrule
SBERT~\cite{reimers2019sentence}  & 10.06\,{\tiny$\pm$0.01} & 30.25\,{\tiny$\pm$0.09} & 0.399 & 1.011 & 0.094 \\
CLAP~\cite{wu2023large}         & 10.02\,{\tiny$\pm$0.04} &29.62\,{\tiny$\pm$0.21} & 0.388 & 1.013 & 0.092 \\
CLIP~\cite{radford2021learning}    & \textbf{9.93\,{\tiny$\pm$0.03}} & \textbf{32.07\,{\tiny$\pm$0.14}} & \textbf{0.413} & \textbf{0.995} & \textbf{0.109} \\
\bottomrule
\end{tabular}
}
\vspace{-7mm}
\end{table}

\vspace{-3mm}
\subsection{Cross-Subject Generalization}
\label{sec:scaling}
\vspace{-2mm}

We evaluated cross-subject generalization in two complementary ways: (1) scaling the number of training subjects, and (2) varying which subjects are held out. Figure~\ref{fig:scaling_curve} shows performance as the number of training subjects $k$ increases, evaluated on a fixed held-out pair (sub-23, sub-24). Our SENSE consistently outperforms both baselines, with the largest gains in the low-data regime. Notably, SENSE already surpasses FE-Phoneme's $k{=}18$ WER at $k{=}2$, and matches or exceeds it on MCD, Mel-Correlation, and STOI by $k{=}8$. Since evaluation is conducted on unseen subjects, these results reflect cross-subject generalization rather than in-distribution scaling, suggesting that SENSE learns representations driven by stimulus content rather than subject-specific patterns.

To test whether SENSE's advantage depends on the choice of held-out subjects, we retrained all models from scratch using two additional held-out pairs, (sub-1, sub-2) and (sub-7, sub-8). Figure~\ref{fig:cross_subject} reports per-subject performance across the resulting six held-out subjects. SENSE outperforms both baselines on every metric and every subject, with consistent relative gains over FE-Phoneme. Although absolute performance varies modestly across subjects, the relative ranking of the methods remains consistent. Full per-subject results are provided in Appendix~\ref{app:per_subject}.

\vspace{-3mm}
\subsection{Ablation Study}
\label{sec:ablations}
\vspace{-2mm}
Table~\ref{tab:ablation} shows that all components of the proposed SENSE contribute to overall performance. Removing ESC degrades semantic metrics, while removing CTC causes an even larger increase in WER, indicating that phoneme-level supervision also preserves intelligibility (Table~\ref{tab:ablation}). Spatial components (GNN, CNN skip, channel attention) provide consistent gains, complementing both acoustic and semantic modeling. Adding ESC alone to baseline encoders improves them but still falls short of SENSE (Table~\ref{tab:esc_ablation}), showing that semantic conditioning is insufficient without our encoder design. To verify that ESC exploits genuine EEG-semantic correspondence rather than acting as generic regularization, we trained SENSE with CLIP targets randomly shuffled within each batch (Table~\ref{tab:ablation}, \emph{w shuffled ESC} row). Shuffled targets yield performance comparable to (and slightly below) the w/o ESC baseline, confirming that ESC's gain depends on correct semantic targets rather than on auxiliary supervision alone.

We compared four graph adjacency schemes (Table~\ref{tab:graph_ablation}). The anatomy-based BioSemi topology outperforms learnable GAT, fully-connected, and random adjacencies, indicating that local distance-based connectivity is the relevant inductive bias. Jointly learning the graph from only 7,200 pairs is under-constrained, while removing structure entirely is more harmful. Adding the CNN skip on top of the BioSemi GNN further improves performance, confirming complementarity between graph-based spatial mixing and channel-wise feature aggregation.

\begin{figure}[t!]
    \centering
    \includegraphics[width=0.9\linewidth]{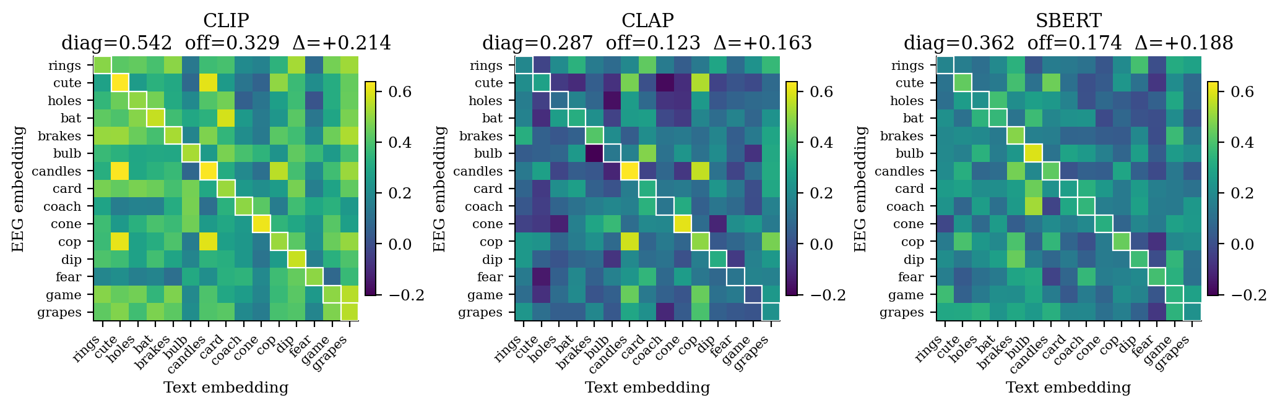}
    \vspace{-4mm}
    \caption{\textbf{EEG–text alignment across semantic encoders.} Cosine similarity matrices between mean EEG embeddings (rows) and text embeddings (columns) for matched words. Stronger word-specific alignment appears as a brighter diagonal. $\Delta$ is the gap between diagonal and off-diagonal means.}
    \label{fig:heatmap}
        \vspace{1mm}
%\end{figure}
% ==== GNN design
%\begin{figure}[t!]
    \centering
    \includegraphics[width=0.9\linewidth]{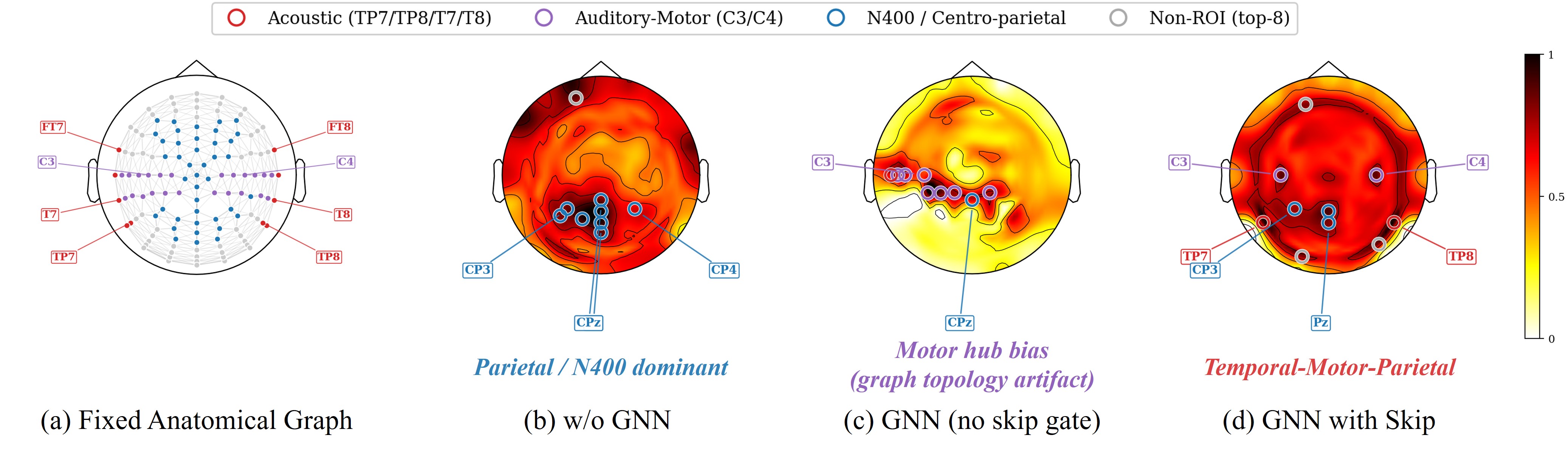}
            \vspace{-3mm}
    \caption{\textbf{ESC attribution across encoder variants.} (a) Anatomical graph used in SENSE. (b--d) Per-channel gradient magnitude of EEG-CLIP alignment, normalized per model. (b) Without GNN: centro-parietal concentration consistent with N400. (c) GNN without skip: bias toward central hub electrodes, a graph-topology artifact. (d) SENSE: tripartite pattern over temporal-auditory (TP7/TP8), motor (C3/C4), and centro-parietal (CPz). Circles: top-8 channels per model; colors: functional ROI.}
    \label{fig:gnn}
                \vspace{-5mm}
\end{figure}

We also evaluated three pretrained semantic encoders for the ESC alignment target (Table~\ref{tab:esc_encoder}): CLIP, CLAP~\cite{wu2023large}, and SBERT~\cite{reimers2019sentence}. CLIP outperforms CLAP and SBERT on all metrics. CLAP, despite being trained for audio--language alignment, performs worse than text-only SBERT, possibly because the N400 stimuli often involve concrete, image-able nouns (\textit{chair}, \textit{dog}, \textit{ants}) for which CLIP's image--text pretraining provides more discriminative semantic structure. Figure~\ref{fig:heatmap} corroborates this: CLIP shows the brightest diagonal and the largest gap between diagonal and off-diagonal values.

%CLIP achieves the best performance on all metrics. Although CLAP is trained for audio–language alignment, it performs slightly worse than the text-only SBERT. One possible explanation is that CLAP is optimized for broad audio-text alignment, whereas the N400 paradigm emphasizes sentence-final lexical semantics. In contrast, CLIP may be better suited to this setting because the N400 stimuli often involve concrete, image-able nouns (\textit{chair}, \textit{dog}, \textit{ants}), for which image-text pretraining provides more discriminative semantic structure. Figure~\ref{fig:heatmap} visualizes cosine similarity between mean EEG embeddings and text embeddings of matched words: CLIP exhibits the brightest diagonal and the largest gap between diagonal and off-diagonal values.

% ==== CLIP CLAP BERT heatmap

\vspace{-4mm}
\subsection{Channel Attribution Analysis}
\label{sec:analysis}
\vspace{-3mm}

We complemented the quantitative results with two analyses of the model's internal behavior, examining whether the learned representations align with our motivation that EEG during speech perception encodes information beyond auditory regions. The channel attention gate (Section~\ref{sec:eeg_module}) is initialized with a positive bias toward auditory-related electrodes and zero elsewhere; no prior is placed on sensorimotor or centro-parietal regions, so the patterns below emerge from training rather than design. The \textit{w/o Ch.Att} row in Table~\ref{tab:ablation}  initializes the gate without the auditory warm-start (all weights set to zero); the moderate performance drop indicates that the gate's emergent non-auditory attribution is not merely an artifact of initialization. After training, auditory-prior channels are modestly down-weighted while central electrodes (C3, C4) and frontal-central regions gain importance, with occipital channels suppressed (Appendix~\ref{app:gate_viz}).

To further investigate which channels contribute to EEG--text alignment, we computed per-channel gradient magnitudes of the ESC alignment (Figure~\ref{fig:gnn}). We treat these gradient maps as model-internal diagnostics rather than evidence of neural sources, given known limitations of saliency-based attribution. The maps reveal three patterns: without a GNN, attribution concentrates over centro-parietal regions consistent with the N400 response. A GNN without a skip connection shifts emphasis to central hub electrodes, a topology artifact of dense connectivity rather than a physiological signal. The full SENSE model recovers a distributed pattern over temporal-auditory (TP7, TP8), central motor (C3, C4), and centro-parietal (CPz) regions. The prominence of C3/C4, which are not explicitly favored by design, is consistent with prior findings that speech perception engages sensorimotor cortex during passive listening~\cite{wilson2004listening,assaneo2018coupling,park2015frontal}; we read this as the model capturing distributed information aligned with known properties of speech processing, not as direct evidence of neural generators. This supports our central hypothesis that EEG-to-speech generation benefits from non-auditory channels. Training dynamics are detailed in Appendix~\ref{app:attribution_trajectory}.

\vspace{-2mm}
\section{Conclusions}
\label{sec:conclusions}
\vspace{-2mm}

We argued that EEG-to-speech should be approached as a joint acoustic-semantic problem, not as acoustic reconstruction alone. We introduced \textbf{SENSE}, which combines a graph-based spatial encoder over electrode geometry with EEG Semantic Conditioning (ESC). On the N400 dataset, SENSE consistently outperforms prior methods on both acoustic and semantic metrics, with particularly strong cross-subject generalization. Channel attribution further recovers patterns spanning auditory, sensorimotor, and centro-parietal regions consistent with neuroscience of speech-perception. However, reliable word-level reconstruction remains challenging, as indicated by WER values near 1, and full waveform synthesis has only been evaluated on the controlled N400 corpus despite promising representation-level adaptation to naturalistic speech. Channel attribution remains model-internal rather than neural-source localization, and evaluation on larger naturalistic and active-speech datasets is an important next step.

\subsection{Broader Impact}
\label{app:broader_impact}
This work uses a previously released public EEG benchmark collected under prior ethical approval, with no new participant recruitment, neural data collection, or sensing hardware, and should not be interpreted as unrestricted thought decoding. Active listening and laboratory conditions are not reliable safeguards, and future applications should protect neural data through measures such as on-device processing, explicit opt-in consent, restrictions on unauthorized access, and deployment-specific oversight and governance. These considerations follow the UNESCO Recommendation on the Ethics of Neurotechnology (2025), including its guidance on protecting neural data and information that may enable inferences about mental states.

\section*{Acknowledgments}
This work was supported in part by the National Research Foundation of Korea(NRF) grant funded by the Korea government(MSIT) (RS-2025-02217071), in part by the Institute of Information \& communications Technology Planning \& Evaluation (IITP) grant funded by the Korea government(MSIT) (RS-2021-II211341), and in part by the AI Seoul Tech Research Support Program of the Seoul Future Foundation.

\clearpage
\bibliography{main}
\bibliographystyle{unsrtnat}

%%%%%%%%%%%%%%%%%%%%%%%%%%%%%%%%%%%%%%%%%%%%%%%%%%%%%%%%%%%%

\appendix
\newpage
 \setcounter{equation}{0}
 \renewcommand{\theequation}{A.\arabic{equation}}
 \setcounter{figure}{0}
 \renewcommand{\thefigure}{A.\arabic{figure}}
 \setcounter{table}{0}
 \renewcommand{\thetable}{A.\arabic{table}}
 
\section{Implementation Details}
\label{app:impl}

This appendix collects the implementation details omitted from Section~\ref{sec:method}, including dataset preprocessing, layer-level architectures, and training configurations. The implementation is provided as supplementary material; pretrained checkpoints will be released with the de-anonymized version upon acceptance.

%------------------------------------------------------------------ 
\subsection{Dataset Preprocessing} 
\label{app:preprocessing} 

\paragraph{EEG Preprocessing.} We follow the preprocessing pipeline of FESDE~\cite{lee2024fesde}. Raw 128-channel EEG signals recorded at 512\,Hz are processed as follows: (1) a 60\,Hz notch filter removes power-line interference; (2) a 0.5--50\,Hz Butterworth bandpass filter isolates the frequency range relevant to speech-related neural activity; (3) Independent Component Analysis (ICA) is applied to remove eye-blink artifacts, with components correlated with EOG channels removed automatically; (4) signals are resampled to 256\,Hz to reduce computational cost while preserving the relevant spectral content. Trials are segmented into fixed windows aligned with sentence onsets and zero-padded to a uniform length when necessary. 

\paragraph{Speech Preprocessing.} The recorded speech audio is downsampled from its native sampling rate to 22{,}050\,Hz to match the input rate of the HiFi-GAN vocoder~\cite{kong2020hifi} used in VITS. Mel-spectrogram targets are extracted with 80 mel bands using a window size of 1024 samples, hop size of 256 samples, and FFT size of 1024, following the VITS configuration~\cite{kim2021conditional}. 

\paragraph{Subject Selection.} Following~\cite{lee2024fesde,lee2025phoneme}, we exclude four subjects (sub-5, sub-10, sub-15, sub-18) due to documented issues with EEG signal quality or behavioral responses, leaving 20 subjects in the analysis. Of these, 18 are used for training and validation, with sub-23 and sub-24 held out for the unseen-subject test conditions.
%------------------------------------------------------------------
\subsection{Structure-Aware EEG Module}
\label{app:eeg_module}

\paragraph{Channel Attention Gate.}
The gate weights $\mathbf{w} \in \mathbb{R}^{128}$ are initialized as $w_i = +2.0$ for the six auditory-related electrodes (T7, T8, FT7, FT8, TP7, TP8), giving $\sigma(2.0) \approx 0.88$ at the start of training, and $w_i = 0.0$ for all remaining channels, giving $\sigma(0) = 0.5$. All gate weights are learnable and updated by gradient descent throughout training.

\paragraph{Graph Adjacency Matrix.}
Electrode positions $p_i \in \mathbb{R}^3$ are taken from the BioSemi 128-channel standard montage available through MNE-Python~\cite{gramfort2013meg}, expressed in meters. The adjacency matrix $\mathbf{A} \in \mathbb{R}^{128 \times 128}$ is constructed as:
\begin{equation}
  A_{ij} = \begin{cases}
    \exp\!\left(-\dfrac{\|p_i - p_j\|_2^2}{2\sigma_p^2}\right)
      & \text{if } \|p_i - p_j\|_2 < \delta, \\
    0 & \text{otherwise,}
  \end{cases}
\end{equation}
with $\sigma_p = 2\,\text{cm}$ and threshold $\delta = 4\,\text{cm}$. The diagonal is set to zero, and the matrix is row-normalized: $\mathbf{A} \leftarrow D^{-1}\mathbf{A}$, where $D$ is the diagonal degree matrix. The resulting adjacency is registered as a fixed buffer and never updated during training.

\paragraph{Graph Convolution Layer.}
At each time step, the scalar activation of each electrode is projected to a $d_{\mathrm{node}} = 64$ dimensional feature via a shared Linear-GELU block. Two layers of residual graph convolution are then applied, where each $f^{(\ell)}$ is a Linear $\to$ LayerNorm $\to$ GELU $\to$ Dropout(0.1) block with hidden dimension $d_{\mathrm{node}} = 64$. After the two message-passing steps, node representations are averaged over all 128 nodes and linearly projected to the latent dimension $d = 192$.

\paragraph{Gated CNN Skip Connection.}
The CNN branch consists of two 1D convolutional layers with kernel size 4, stride 1, and padding 3, mapping from 128 input channels to $d = 192$ output channels. Each convolutional layer is followed by GELU activation. The branch output is normalized to match the GNN output's empirical standard deviation:
\begin{equation}
  \tilde{\mathbf{H}}_{\mathrm{CNN}} =
    \frac{\mathrm{std}(\mathbf{H}_{\mathrm{GNN}})}{\mathrm{std}(\mathbf{H}_{\mathrm{CNN}})}
    \,\mathbf{H}_{\mathrm{CNN}},
\end{equation}
where $\mathrm{std}(\cdot)$ is computed over the channel and time dimensions. The scalar mixing gate $g \in \mathbb{R}$ is initialized to $-2.0$, giving $\sigma(-2) \approx 0.12$ at the start of training. The gate is fully learnable and adapts during optimization.

\paragraph{Temporal Encoding.}
A single Conv1d layer with kernel size 3, stride 3, and padding 1 reduces the temporal resolution by a factor of 3, followed by four S4 blocks~\cite{gu2021efficiently} with hidden dimension $d = 192$, state dimension 64, and dropout 0.1. Each S4 block follows the standard configuration of \cite{gu2021efficiently} with bidirectional state-space mixing.

\paragraph{EEG Decoder (training only).}
The EEG decoder consists of a single ConvTranspose1d block with kernel size 3, stride 3, and padding 0 for temporal upsampling, followed by a two-layer ConvTranspose1d stack with kernel size 4, stride 1, and padding 3 that restores the channel dimension from $d = 192$ back to 128. The reconstruction loss is the cosine similarity loss of FESDE~\cite{lee2024fesde}, applied with length masking that excludes zero-padded frames. The decoder is discarded at inference.

\paragraph{Connector.}
The connector is a 6-layer multi-head self-attention encoder with hidden dimension 192, 2 attention heads, feed-forward dimension 768, kernel size 3, and dropout 0.1, following the attention encoder of VITS~\cite{kim2021conditional}. The output is projected by a $1\times 1$ convolution to the mean and log-variance parameters of the VITS prior. A stop-gradient is applied between $\mathbf{M}$ and the connector input, isolating the EEG Module from the speech-pathway gradients.

%------------------------------------------------------------------
\subsection{Phoneme Sequence Supervision}
\label{app:ctc}

%The phoneme predictor is taken from FE-Phoneme~\cite{lee2025phoneme} without modification. It consists of one Conformer block~\cite{gulati2020conformer} with encoder dimension 192, 8 attention heads, feed-forward expansion factor 4, convolutional kernel size 31, and dropout 0.1, followed by an LSTM attention decoder with hidden dimension 192, location-based attention, and a maximum decode length of 50 over a 51-symbol IPA vocabulary (45 IPA phonemes, 3 punctuation symbols, and pad/SOS/EOS tokens). The CTC loss~\cite{graves2006connectionist} is applied to the decoder output with $\mathrm{zero\_infinity}=\mathrm{True}$ and $\mathrm{blank}=0$. The teacher forcing ratio is set to 0 throughout training, so the decoder is fully autoregressive on its own predictions. The predictor is discarded at inference. We refer the reader to \cite{lee2025phoneme} for full architectural details.

The phoneme predictor is taken from FE-Phoneme~\cite{lee2025phoneme} without modification. It consists of one Conformer block~\cite{gulati2020conformer} with encoder dimension 192, 8 attention heads, feed-forward expansion factor 4, convolutional kernel size 31, and dropout 0.1, followed by an LSTM attention decoder with hidden dimension 192, location-based attention, and a maximum decode length of 50 over a 51-symbol IPA vocabulary (45 IPA phonemes, 3 punctuation symbols, and pad/SOS/EOS tokens). The teacher forcing ratio is set to 0 throughout training, so the decoder is fully autoregressive on its own predictions.

\paragraph{CTC objective.}
Given a target phoneme sequence $\mathbf{z} = (z_1, \ldots, z_L)$ of length $L$, the LSTM attention decoder produces a sequence of $L$ logit vectors $\mathbf{Y} = (\mathbf{y}_1, \ldots, \mathbf{y}_L)$ over the 51-symbol vocabulary (including a blank token). The CTC objective~\cite{graves2006connectionist} marginalizes over alignments $\boldsymbol{\pi}$ that map to $\mathbf{z}$ under the standard CTC reduction $B$:
\begin{equation}
\mathcal{L}_{\mathrm{ctc}} =
    -\log\!\sum_{\boldsymbol{\pi} \in B^{-1}(\mathbf{z})}
    P(\boldsymbol{\pi} \mid \mathbf{Y}).
\end{equation}
We use the PyTorch \texttt{CTCLoss} with $\mathrm{zero\_infinity}{=}\mathrm{True}$ and $\mathrm{blank}{=}0$, and set both $\mathrm{input\_lengths}$ and $\mathrm{target\_lengths}$ to the actual phoneme sequence length (excluding the SOS token) per sample, following the FE-Phoneme implementation~\cite{lee2025phoneme}. The predictor is discarded at inference.

%------------------------------------------------------------------
\subsection{EEG Semantic Conditioning}
\label{app:esc}

\paragraph{CLIP Embeddings.}
We use frozen CLIP ViT-B/32~\cite{radford2021learning} to compute text embeddings of dimension 512. Embeddings for all sentences in the dataset are precomputed once and cached to disk. The CLIP encoder itself is not loaded during training.

\paragraph{Projection layers.}
The decoder-side projection $W_{\mathrm{clip}} \colon \mathbb{R}^{512} \to \mathbb{R}^{192}$ is a single \texttt{nn.Linear} layer that is trained jointly with the speech decoder under the acoustic objectives (mel reconstruction, KL, adversarial, feature-matching). The EEG-side projection $W_{\mathrm{eeg}} \in \mathbb{R}^{192 \times 192}$ is a single \texttt{nn.Linear} layer trained only by $\mathcal{L}_{\mathrm{esc}}$; gradients from the conditioning bridge path are blocked at $W_{\mathrm{eeg}}$ so that the only gradient signal reaching $W_{\mathrm{eeg}}$ is the ESC alignment loss.

\paragraph{Length-masked temporal mean.}
$\mathbf{g}_{\mathrm{eeg}}$ is computed by first taking a length-masked temporal mean of $\mathbf{M}$ over its time dimension, where the mask excludes zero-padded frames according to each sample's actual EEG length. The masked mean is then projected by $W_{\mathrm{eeg}}$ to produce $\mathbf{g}_{\mathrm{eeg}} \in \mathbb{R}^{192}$.

\paragraph{Conditioning bridge schedule.}
At each training step, with probability $p_{\mathrm{eeg}} = 0.33$ the decoder's conditioning is set to $\mathbf{g}_{\mathrm{eeg}}$; gradients along this path are blocked at both $\mathbf{M}$ (so the EEG Module is not affected by the conditioning bridge) and $W_{\mathrm{eeg}}$ (so $W_{\mathrm{eeg}}$ is shaped only by $\mathcal{L}_{\mathrm{esc}}$). With the remaining probability $1 - p_{\mathrm{eeg}}$, standard Classifier-Free Guidance dropout~\cite{ho2022classifier} is applied: a per-sample Bernoulli mask with rate $0.5$ selects between the CLIP text conditioning $W_{\mathrm{clip}}(\mathbf{c}_{\mathrm{text}})$ and a learnable null embedding (a single $\mathbb{R}^{192}$ parameter, initialized to zero). The decoder thus encounters three conditioning sources during training, with overall frequencies of approximately 33\%, 33.5\%, and 33.5\% for $\mathbf{g}_{\mathrm{eeg}}$, the CLIP text embedding, and the null embedding, respectively.

\paragraph{Inference.}
At inference, $\mathbf{g}_{\mathrm{eeg}}$ is computed from $\mathbf{M}$ without any text input and replaces the null embedding as the decoder's conditioning signal. The CFG scale is set to 0, so the decoder produces speech conditioned solely on $\mathbf{g}_{\mathrm{eeg}}$ without any text-guidance interpolation.

%------------------------------------------------------------------
\subsection{Training Configuration}
\label{app:training}

\paragraph{Optimizer.}
Following the convention of FESDE~\cite{lee2024fesde} and VITS~\cite{kim2021conditional}, we use two separate optimizers, both AdamW~\cite{loshchilov2017decoupled}. The first optimizer trains the EEG Module (Channel Attention Gate, GNN, CNN skip, Temporal Encoding, EEG Decoder) and the phoneme predictor. The second optimizer trains the speech pathway (Connector, $W_{\mathrm{eeg}}$, $W_{\mathrm{clip}}$, VITS prior, HiFi-GAN generator, and discriminators). The two optimizers are connected through the stop-gradient described in Section~\ref{sec:eeg_module}: $\mathbf{M}$ is computed by the first optimizer's parameters and treated as a fixed input to the second. Both optimizers use learning rate $2 \times 10^{-4}$, $\beta_1 = 0.8$, $\beta_2 = 0.99$, weight decay $0.01$, and exponential learning rate decay with $\gamma = 0.999875$ per epoch.

\paragraph{Loss weights.}
The loss-weight values used in Eq.~\ref{eq:total_loss} are summarized in Table~\ref{tab:lambdas}.

\begin{table}[h]
\centering
\caption{\textbf{Loss weights used in the total training objective.}}
\label{tab:lambdas}
\setlength{\tabcolsep}{40pt}
\begin{tabular}{lll}
\toprule
Symbol & Value & Source \\
\midrule
$\lambda_{\mathrm{mel}}$ & 45 & VITS default \\
$\lambda_{\mathrm{kl}}$ & 1.0 & VITS default \\
$\lambda_{\mathrm{recon}}$ & 1.0 & FESDE default \\
$\lambda_{\mathrm{ctc}}$ & 0.3 & FE-Phoneme default \\
$\lambda_{\mathrm{esc}}$ & 0.5 & SENSE \\
\bottomrule
\end{tabular}
\end{table}

The adversarial loss $\mathcal{L}_{\mathrm{adv}}$ and feature-matching loss $\mathcal{L}_{\mathrm{fm}}$ follow the VITS implementation exactly~\cite{kim2021conditional} and use the standard HiFi-GAN multi-period and multi-scale discriminators~\cite{kong2020hifi}.

\paragraph{Training duration and hardware.}
\label{app:cost}
SENSE is trained for 900 epochs with 7{,}200 training pairs and batch size 16 per GPU, corresponding to approximately 200{,}000 gradient steps. Training uses full FP32 precision (no automatic mixed precision) on two NVIDIA RTX 3090 GPUs with PyTorch DistributedDataParallel and NCCL backend. We save a checkpoint every 5{,}000 steps and evaluate the checkpoint at step 200{,}000, where validation performance plateaus. Total wall-clock training time for SENSE is approximately 2 days. Baselines (FESDE, FE-Phoneme) are trained for 100{,}000 iterations as in their original papers; we verified that training them longer does not improve performance, as their losses plateau before this point.

\noindent\textbf{Matched Training-Budget Comparison.}
To control for the different training schedules used in the main experiments, we additionally compare all models after 100k training iterations on the unseen-subject split. As shown in Table~\ref{tab:budget_100k}, SENSE remains better than both baselines across all four reconstruction metrics under the same training budget.

\begin{table}[h]
\centering
\caption{\textbf{Matched training-budget comparison} on the unseen-subject split. All models are evaluated after 100k training iterations.}
\label{tab:budget_100k}
\setlength{\tabcolsep}{25pt}
\renewcommand{\arraystretch}{0.8}
\scriptsize{
\begin{tabular}{lcccc}
\toprule
\textbf{Model}
& \textbf{MCD}$\downarrow$
& \textbf{Mel-Corr}$\uparrow$
& \textbf{STOI}$\uparrow$
& \textbf{WER}$\downarrow$ \\
\midrule
FESDE~\cite{lee2024fesde}
& 11.73 & 18.70 & 0.274 & 1.217 \\
FE-Phoneme~\cite{lee2025phoneme}
& 10.65 & 26.90 & 0.381 & 1.123 \\
\textbf{SENSE}
& \textbf{10.24} & \textbf{31.00} & \textbf{0.409} & \textbf{1.033} \\
\bottomrule
\end{tabular}
}
\end{table}

\paragraph{Cross-subject scaling experiments.}
For Section~\ref{sec:scaling}, we use a nested subset structure $k{=}2 \subset k{=}4 \subset k{=}8 \subset k{=}18$. The $k{=}2$ subset is sub-21 and sub-22; $k{=}4$ adds sub-08 and sub-16; $k{=}8$ adds sub-04, sub-09, sub-12, and sub-13; and $k{=}18$ uses all 18 training subjects. The held-out test subjects sub-23 and sub-24 are excluded from every training subset. All scaling experiments use the same training configuration as the main experiments, with the same number of epochs and the same checkpoint selection protocol.

\paragraph{Reproducibility.}
All models are trained with five random seeds. For each seed, the random number generators of PyTorch, NumPy, and Python are initialized identically; CUDA's deterministic mode is enabled. Reported numbers are the mean across seeds, with the standard deviation reported when shown.

%------------------------------------------------------------------
\subsection{Evaluation Protocol}
\label{app:eval}

\paragraph{Whisper transcription.}
We use Whisper-large-v3~\cite{radford2023robust} with the following decoding settings: temperature 0, beam size 5, no condition on previous text. Transcriptions are normalized for WER computation by lowercasing, removing punctuation other than apostrophes, and collapsing repeated whitespace. WER is computed via the \texttt{jiwer} library against the normalized ground-truth transcript.

\paragraph{BERTScore.}
BERTScore~\cite{zhang2019bertscore} is computed using the \texttt{roberta-large} model with rescaling against a baseline. We report the F1 component.

\paragraph{CLIP-Sim.}
We use CLIP ViT-B/32~\cite{radford2021learning} to compute text embeddings of both the Whisper transcription and the ground-truth sentence, and report the cosine similarity between them.

\paragraph{MCD and Mel-Correlation.}  
MCD is computed over the first 25 mel-cepstral coefficients (excluding the 0th coefficient) with dynamic time warping, and Mel-Correlation is the Pearson correlation between mel-spectrograms of generated and ground-truth speech aligned by length truncation.

\paragraph{STOI.}
We use the standard implementation of \cite{taal2010short} from the \texttt{pystoi} library at 16\,kHz, with both signals resampled from 22{,}050\,Hz.

%------------------------------------------------------------------
\subsection{Symbol Reference}
\label{app:symbols}

For convenience, Table~\ref{tab:symbols} summarizes the symbols introduced in Section~\ref{sec:method}.

\begin{table}[h]
\centering
\caption{\textbf{Notation used throughout the method.}}
\label{tab:symbols}
\setlength{\tabcolsep}{20pt}
\begin{tabular}{ll}
\toprule
Symbol & Meaning \\
\midrule
$\mathbf{X} \in \mathbb{R}^{128 \times T}$ & Raw EEG input ($T$ time steps, 128 channels) \\
$\mathbf{X}_{\mathrm{att}}$ & Channel-gated EEG \\
$\mathbf{w} \in \mathbb{R}^{128}$ & Channel attention gate weights (learnable) \\
$\mathbf{A} \in \mathbb{R}^{128 \times 128}$ & Fixed graph adjacency matrix \\
$\mathbf{H}_{\mathrm{GNN}}, \mathbf{H}_{\mathrm{CNN}} \in \mathbb{R}^{d \times T}$ & GNN and CNN branch outputs \\
$\mathbf{H} \in \mathbb{R}^{d \times T}$ & Combined branch output\\
$\mathbf{M} \in \mathbb{R}^{d \times T'}$ & EEG latent ($d = 192$, $T'$ after temporal compression) \\
$d$ & EEG latent dimension ($d = 192$) \\
$d_{\mathrm{node}}$ & Per-node feature dimension in graph conv ($d_{\mathrm{node}} = 64$) \\
$\mathbf{c}_{\mathrm{text}} \in \mathbb{R}^{512}$ & CLIP text embedding (frozen ViT-B/32) \\
$W_{\mathrm{clip}}$ & Decoder-side CLIP projection (learned, dim $512 \to d$) \\
$W_{\mathrm{eeg}}$ & EEG-side projection (learned, dim $d \to d$) \\
$\mathbf{g}_{\mathrm{eeg}}, \mathbf{g}_{\mathrm{clip}} \in \mathbb{R}^{d}$ & EEG and CLIP conditioning vectors \\
$\mathcal{C}$ & Subset of congruent indices in a batch \\
$\mathrm{sg}[\cdot]$ & Stop-gradient \\
$p_{\mathrm{eeg}}$ & Probability of using $\mathbf{g}_{\mathrm{eeg}}$ during training ($p_{\mathrm{eeg}} = 0.33$) \\
\bottomrule
\end{tabular}
\end{table}

\subsection{Additional Qualitative Results}
\label{app:additional_samples}

\begin{figure}[t]
    \centering
    \begin{minipage}[c]{0.05\linewidth}
        \centering
        \vspace{0.0cm}
        \rotatebox{90}{\makebox[2.4cm][c]{\small GT}} \\[0pt]
        \rotatebox{90}{\makebox[2.4cm][c]{\small \textbf{SENSE}}} \\[0pt]
        \rotatebox{90}{\makebox[2.4cm][c]{\small FESDE}} \\[0pt]
        \rotatebox{90}{\makebox[2.4cm][c]{\small FE\_Phoneme}}
    \end{minipage}%
    \begin{minipage}[c]{0.45\linewidth}
        \includegraphics[width=\linewidth]{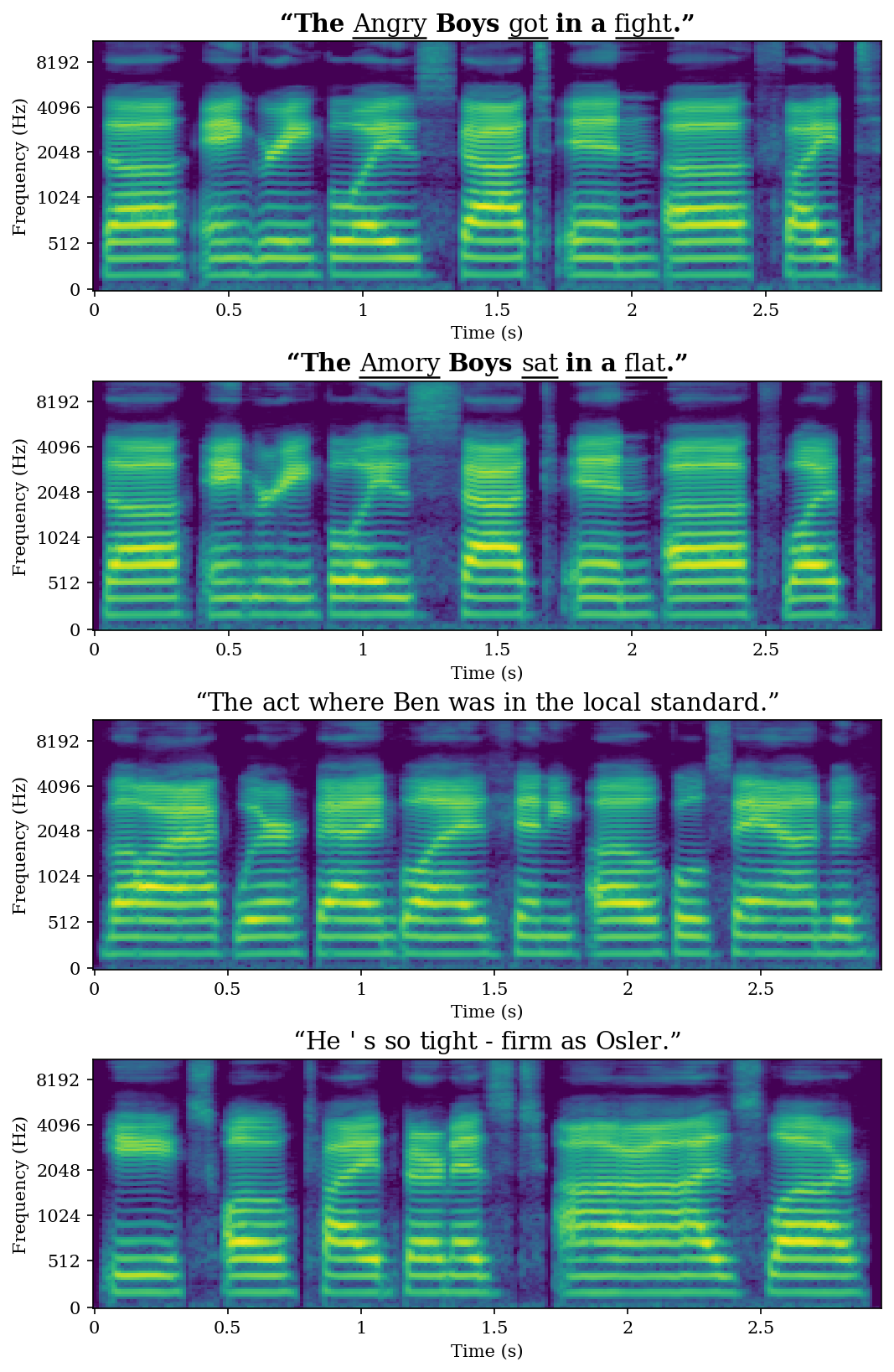}
    \end{minipage}
    \begin{minipage}[c]{0.45\linewidth}
        \includegraphics[width=\linewidth]{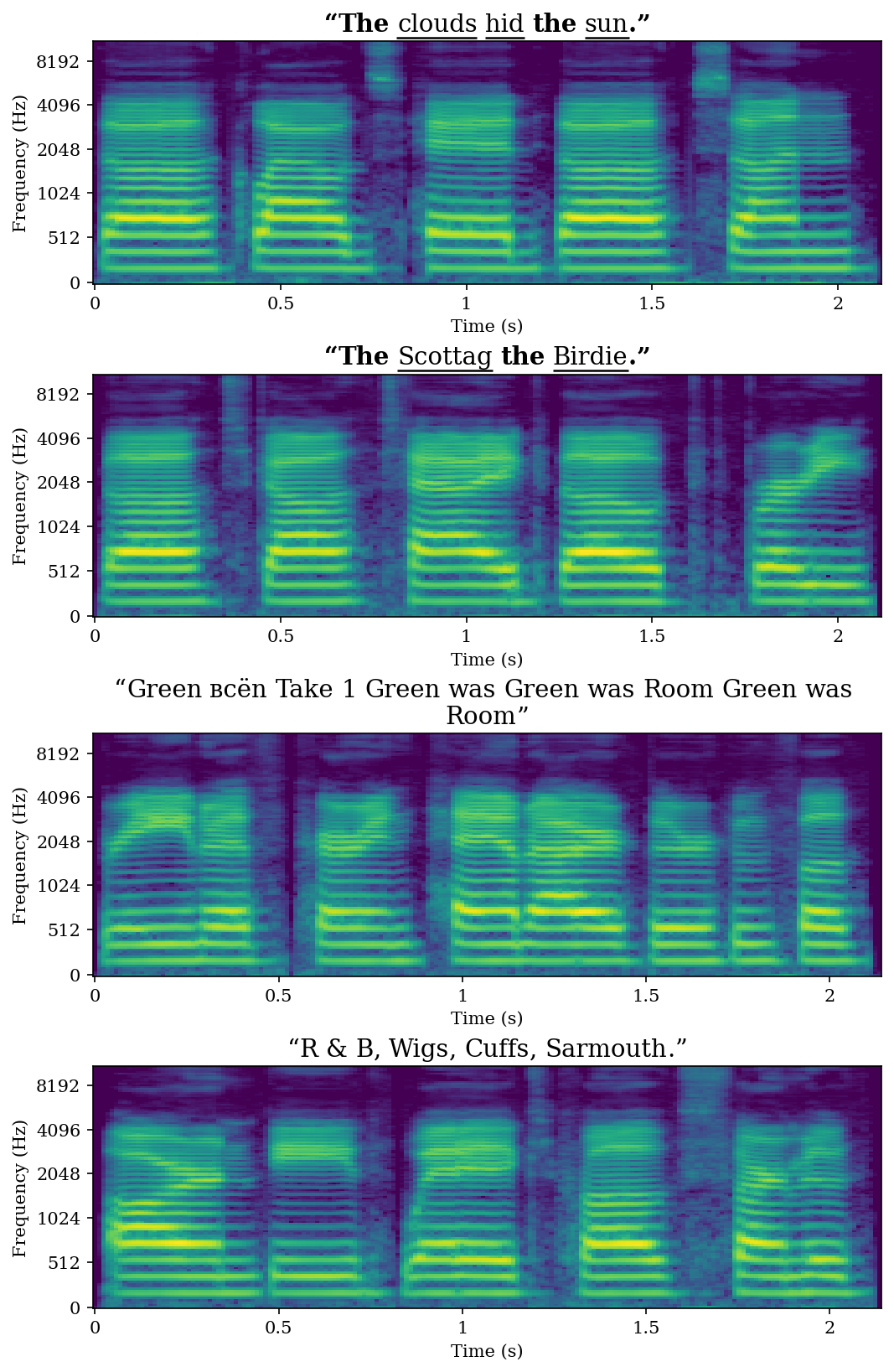}
    \end{minipage}

    \begin{minipage}[c]{0.05\linewidth}
        \centering
        \rotatebox{90}{\makebox[2.4cm][c]{\small GT}} \\[0pt]
        \rotatebox{90}{\makebox[2.4cm][c]{\small \textbf{SENSE}}} \\[0pt]
        \rotatebox{90}{\makebox[2.4cm][c]{\small FESDE}} \\[0pt]
        \rotatebox{90}{\makebox[2.4cm][c]{\small FE\_Phoneme}}
    \end{minipage}%
    \begin{minipage}[c]{0.45\linewidth}
        \includegraphics[width=\linewidth]{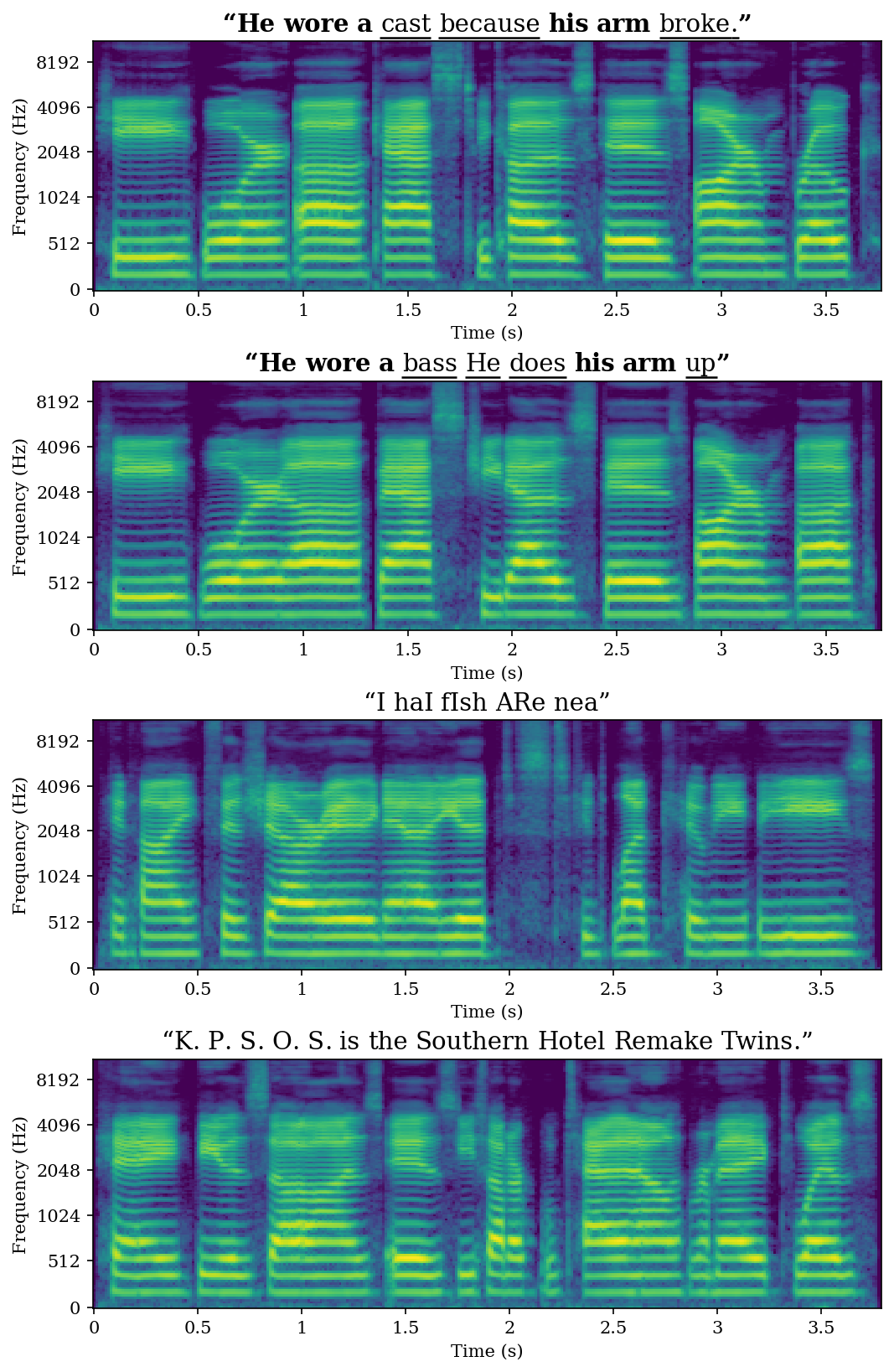}
    \end{minipage}%
    \begin{minipage}[c]{0.45\linewidth}
        \includegraphics[width=\linewidth]{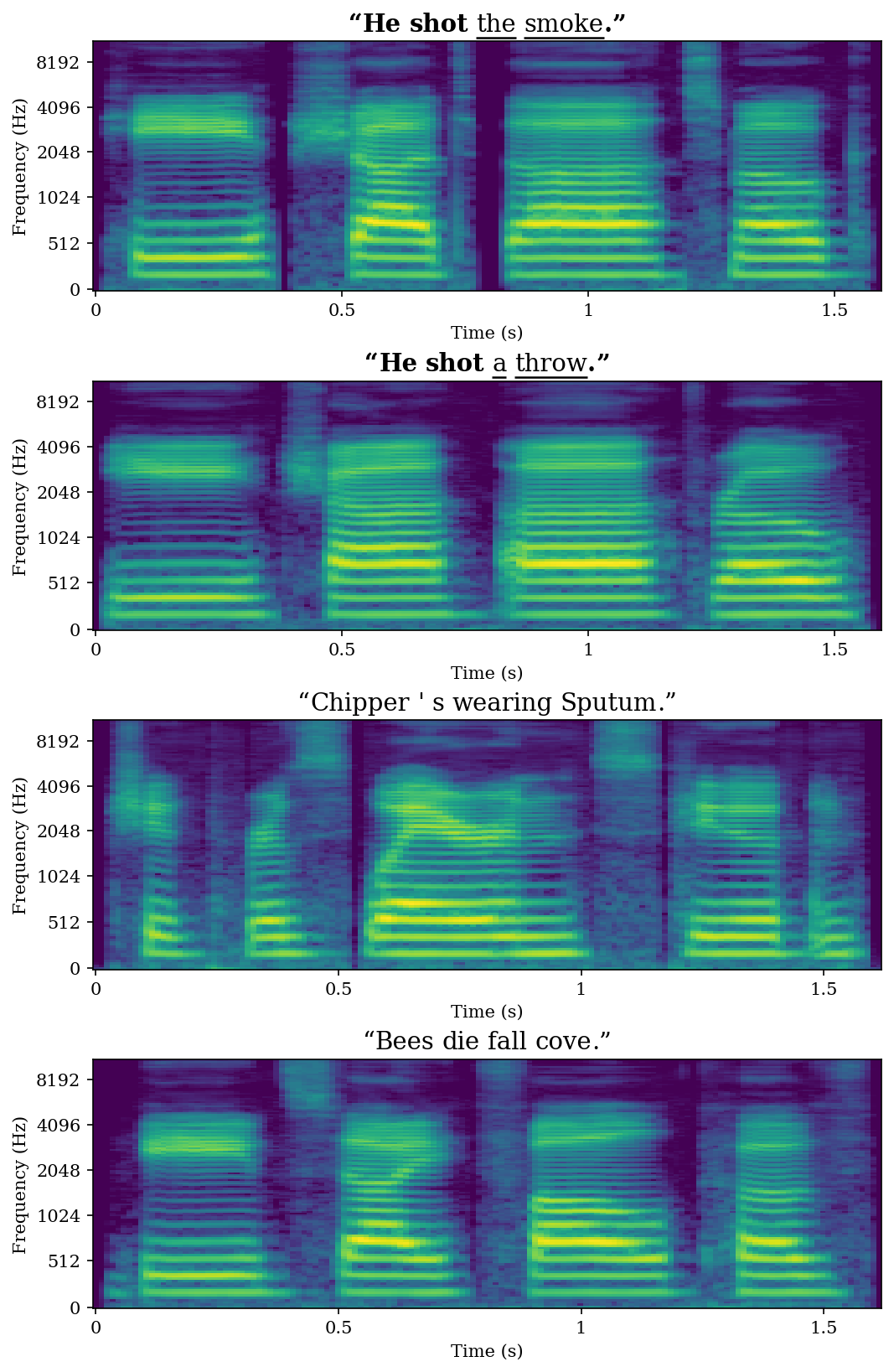}
    \end{minipage}
    
    \caption{Additional Melspectrogram Comparison}
    \label{fig:spectrogram_supp_1}
\end{figure}

\begin{figure}[t]
    \centering
    \begin{minipage}[c]{0.05\linewidth}
        \centering
        \vspace{0.0cm}
        \rotatebox{90}{\makebox[2.4cm][c]{\small GT}} \\[0pt]
        \rotatebox{90}{\makebox[2.4cm][c]{\small \textbf{SENSE}}} \\[0pt]
        \rotatebox{90}{\makebox[2.4cm][c]{\small FESDE}} \\[0pt]
        \rotatebox{90}{\makebox[2.4cm][c]{\small FE\_Phoneme}}
    \end{minipage}%
    \begin{minipage}[c]{0.45\linewidth}
        \includegraphics[width=\linewidth]{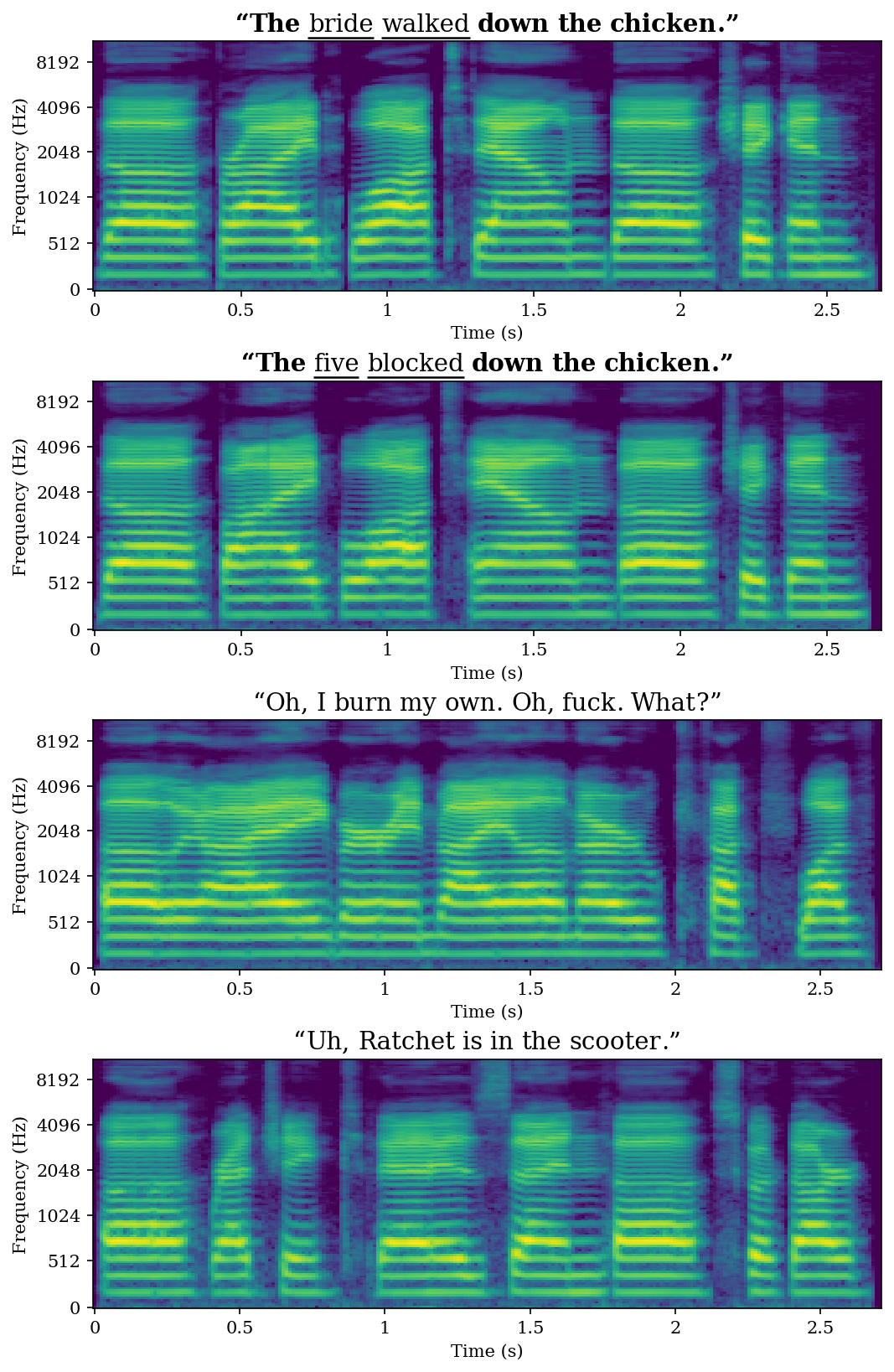}
    \end{minipage}
    \begin{minipage}[c]{0.45\linewidth}
        \includegraphics[width=\linewidth]{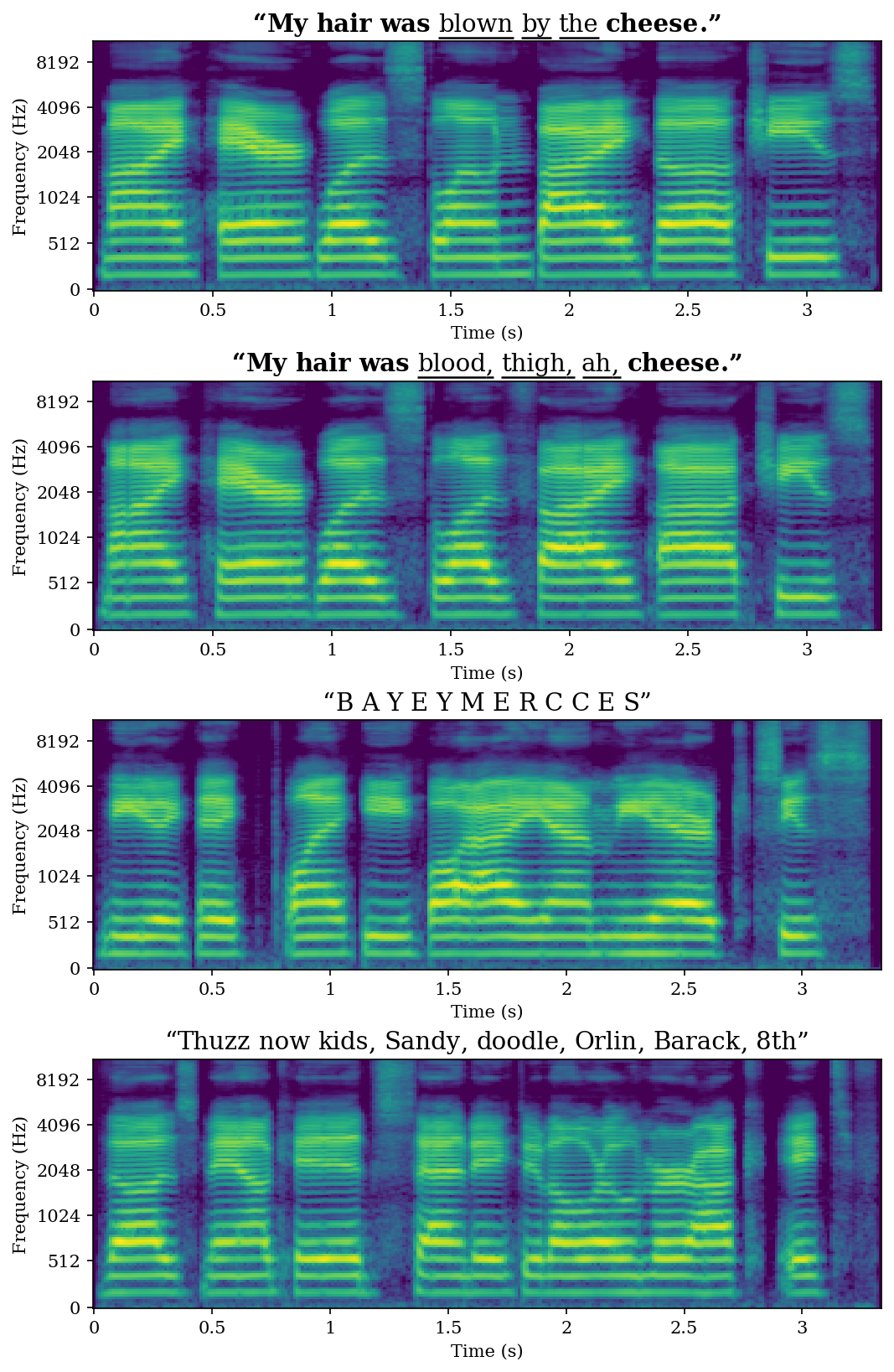}
    \end{minipage}

    \begin{minipage}[c]{0.05\linewidth}
        \centering
        \rotatebox{90}{\makebox[2.4cm][c]{\small GT}} \\[0pt]
        \rotatebox{90}{\makebox[2.4cm][c]{\small \textbf{SENSE}}} \\[0pt]
        \rotatebox{90}{\makebox[2.4cm][c]{\small FESDE}} \\[0pt]
        \rotatebox{90}{\makebox[2.4cm][c]{\small FE\_Phoneme}}
    \end{minipage}%
    \begin{minipage}[c]{0.45\linewidth}
        \includegraphics[width=\linewidth]{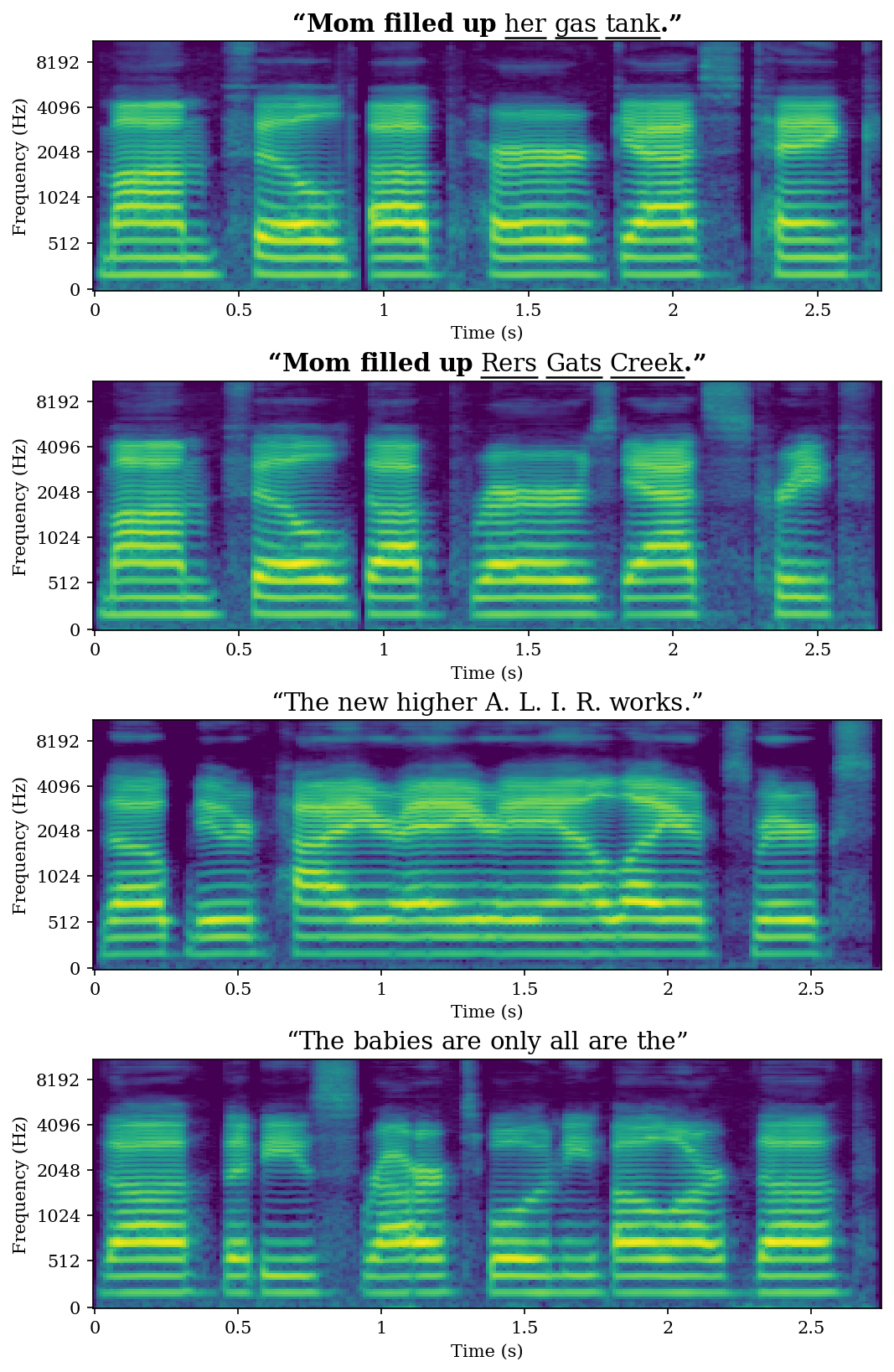}
    \end{minipage}%
    \begin{minipage}[c]{0.45\linewidth}
        \includegraphics[width=\linewidth]{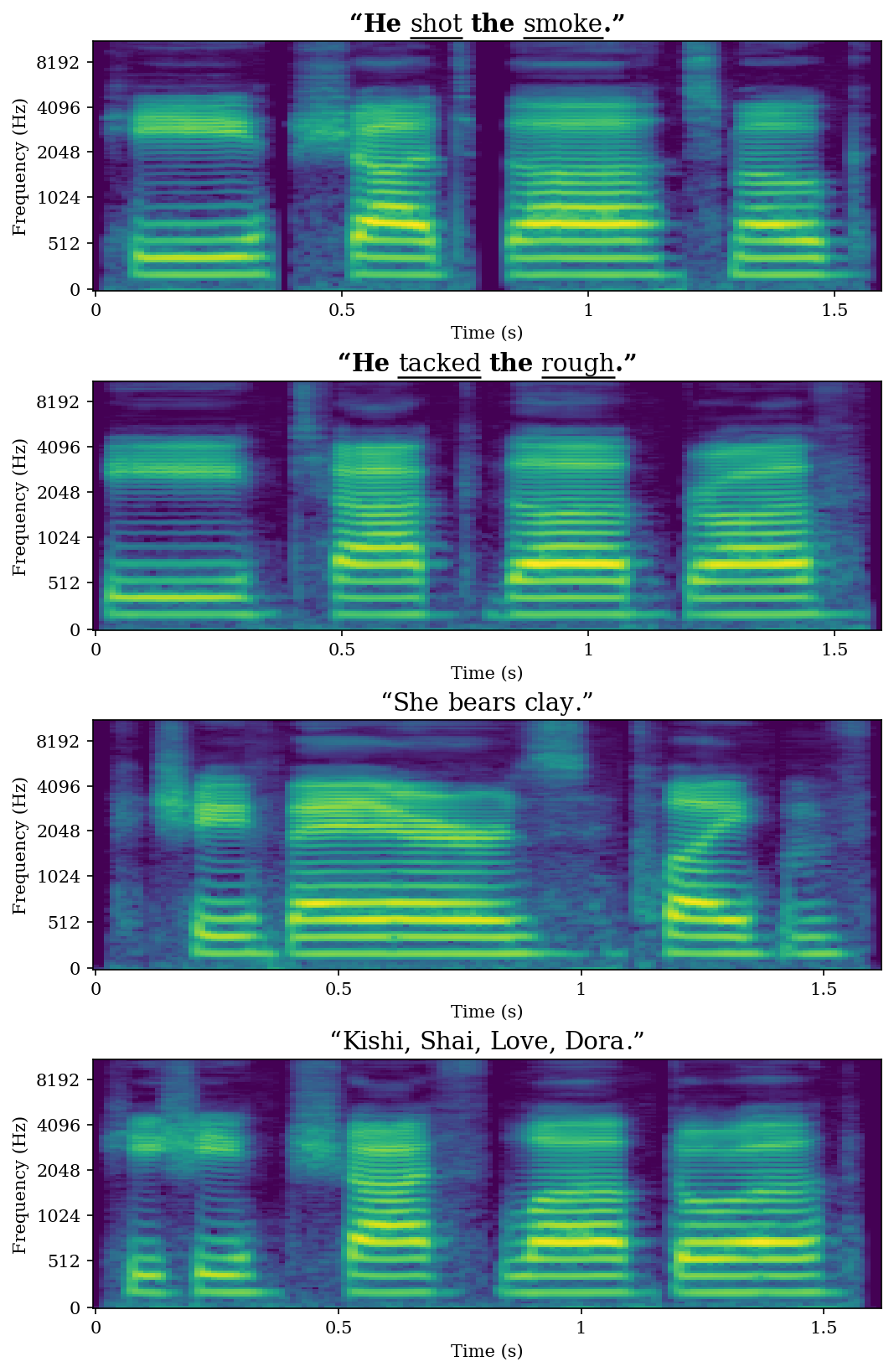}
    \end{minipage}
    
    \caption{Additional Melspectrogram Comparison}
    \label{fig:spectrogram_supp_2}
\end{figure}

\begin{figure}[t]
    \centering
    \begin{minipage}[c]{0.05\linewidth}
        \centering
        \vspace{0.0cm}
        \rotatebox{90}{\makebox[2.4cm][c]{\small GT}} \\[0pt]
        \rotatebox{90}{\makebox[2.4cm][c]{\small \textbf{SENSE}}} \\[0pt]
        \rotatebox{90}{\makebox[2.4cm][c]{\small FESDE}} \\[0pt]
        \rotatebox{90}{\makebox[2.4cm][c]{\small FE\_Phoneme}}
    \end{minipage}%
    \begin{minipage}[c]{0.45\linewidth}
        \includegraphics[width=\linewidth]{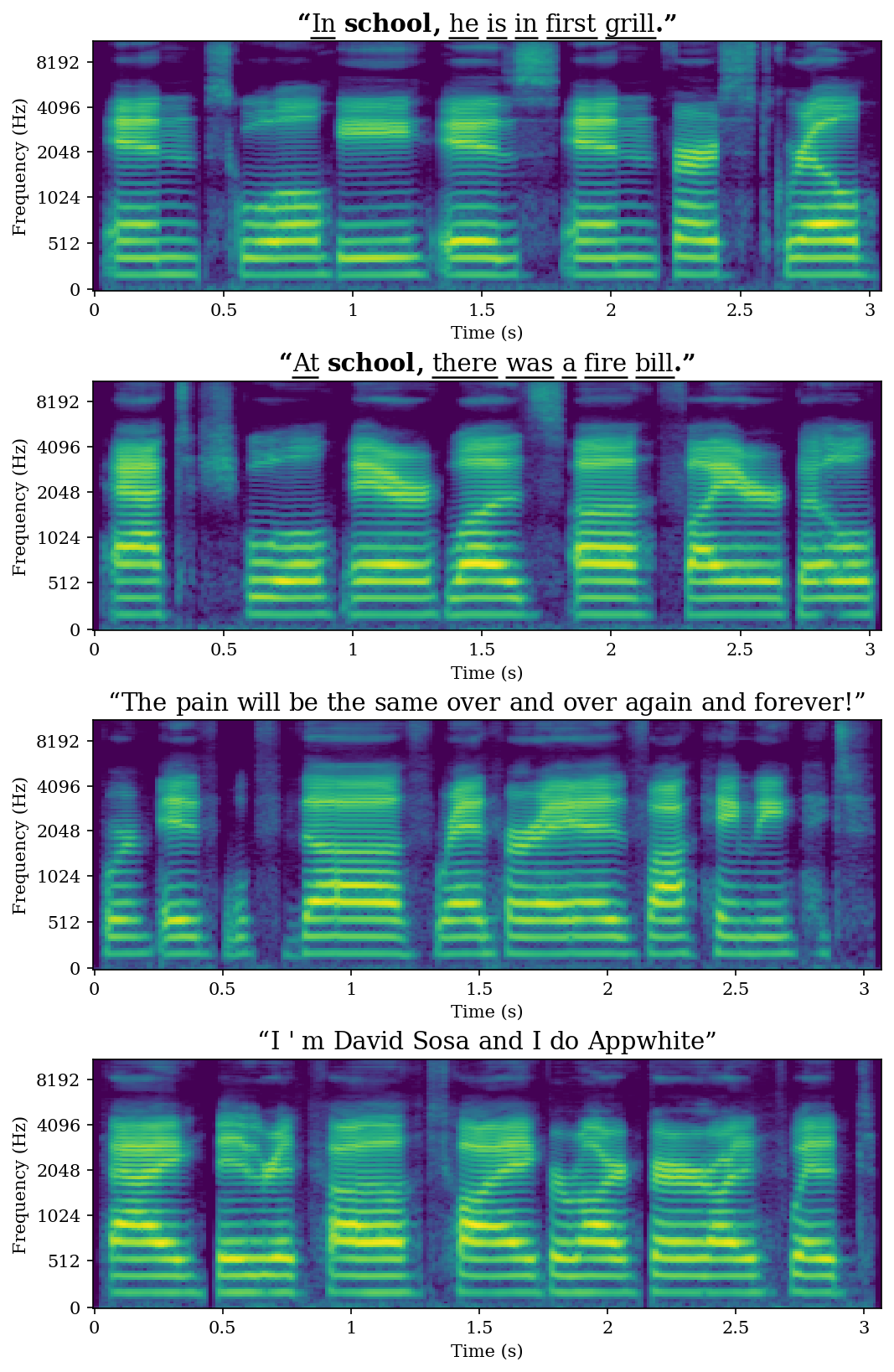}
    \end{minipage}
    \begin{minipage}[c]{0.45\linewidth}
        \includegraphics[width=\linewidth]{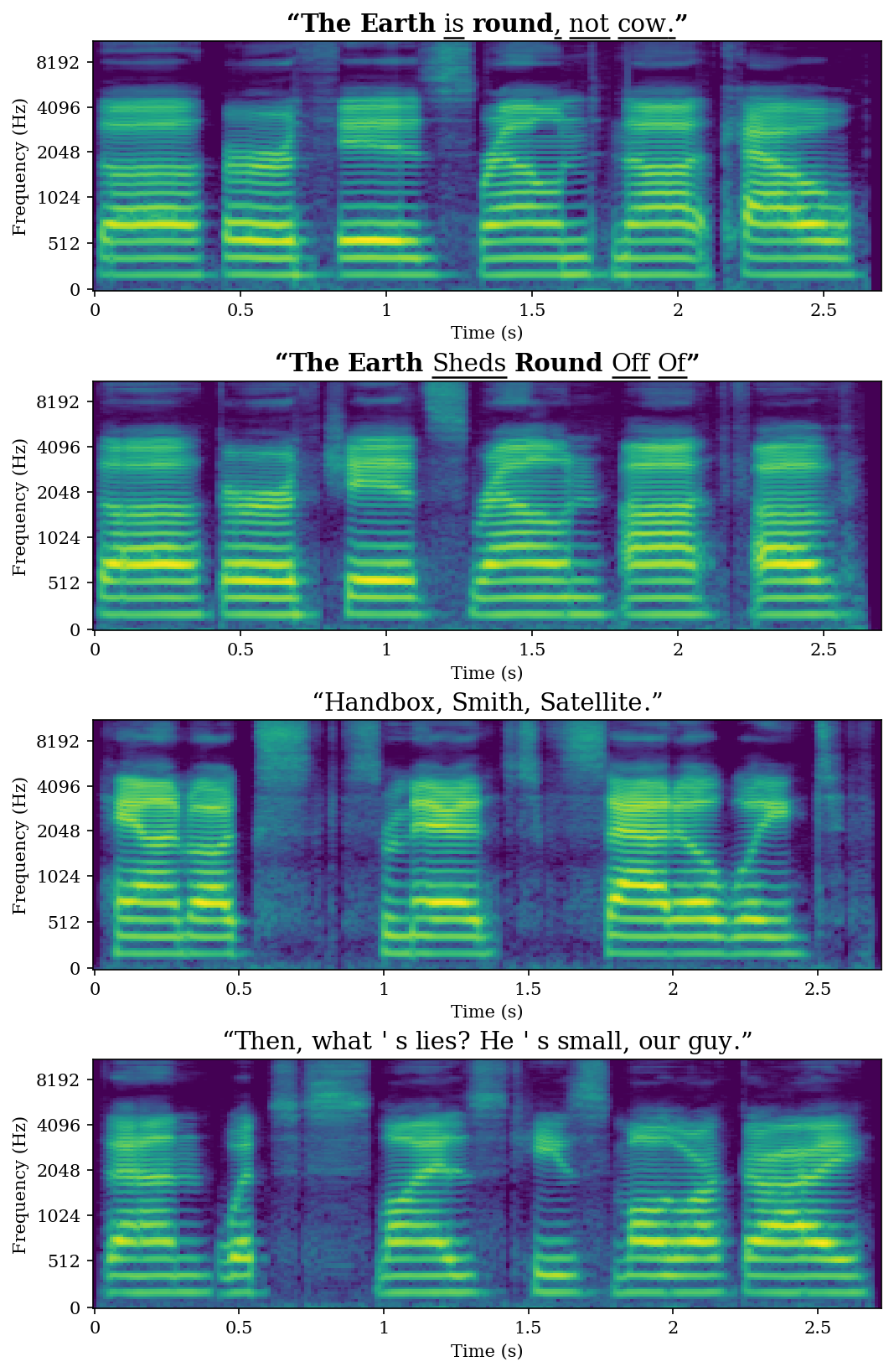}
    \end{minipage}

    \begin{minipage}[c]{0.05\linewidth}
        \centering
        \rotatebox{90}{\makebox[2.4cm][c]{\small GT}} \\[0pt]
        \rotatebox{90}{\makebox[2.4cm][c]{\small \textbf{SENSE}}} \\[0pt]
        \rotatebox{90}{\makebox[2.4cm][c]{\small FESDE}} \\[0pt]
        \rotatebox{90}{\makebox[2.4cm][c]{\small FE\_Phoneme}}
    \end{minipage}%
    \begin{minipage}[c]{0.45\linewidth}
        \includegraphics[width=\linewidth]{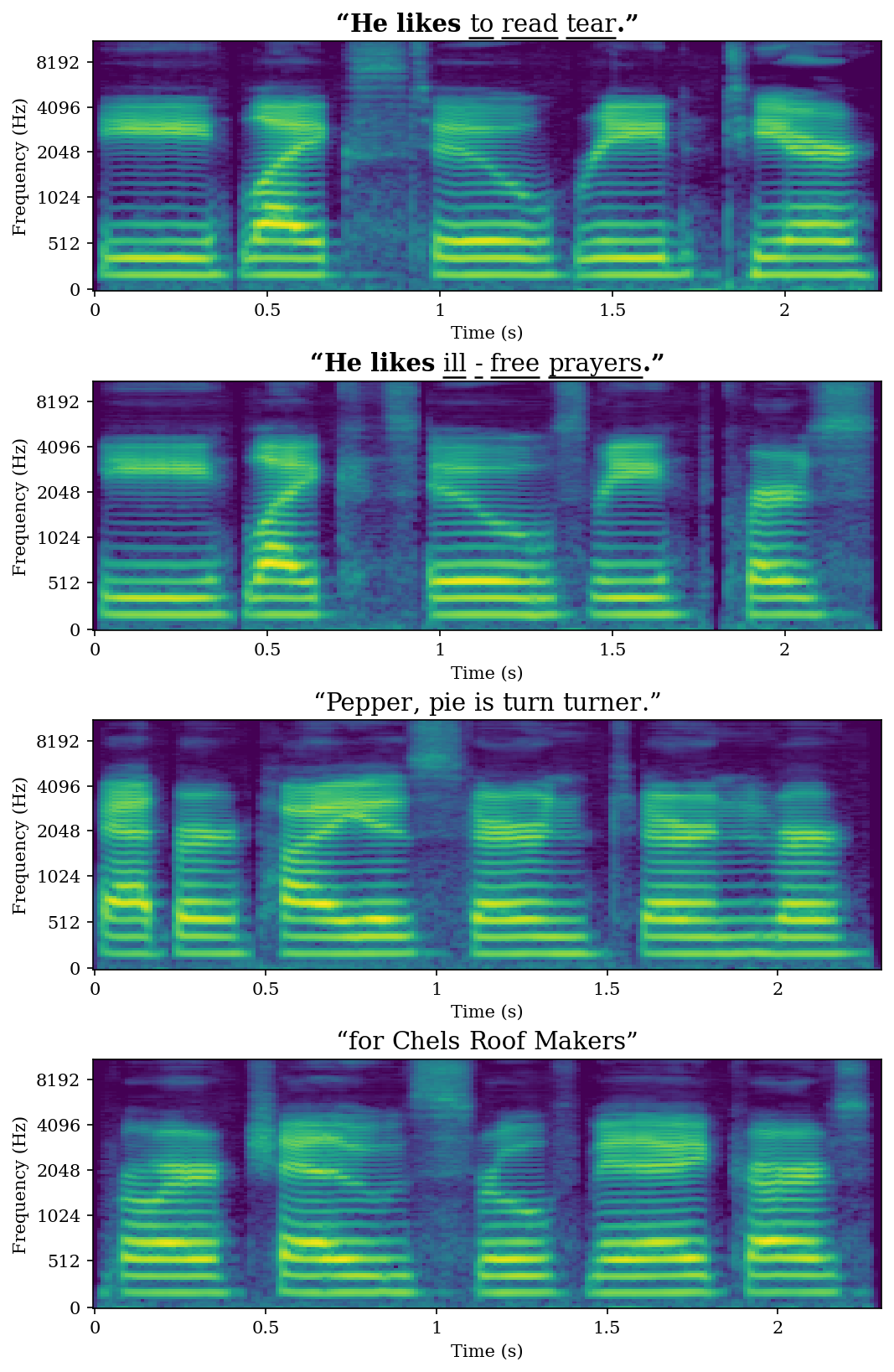}
    \end{minipage}%
    \begin{minipage}[c]{0.45\linewidth}
        \includegraphics[width=\linewidth]{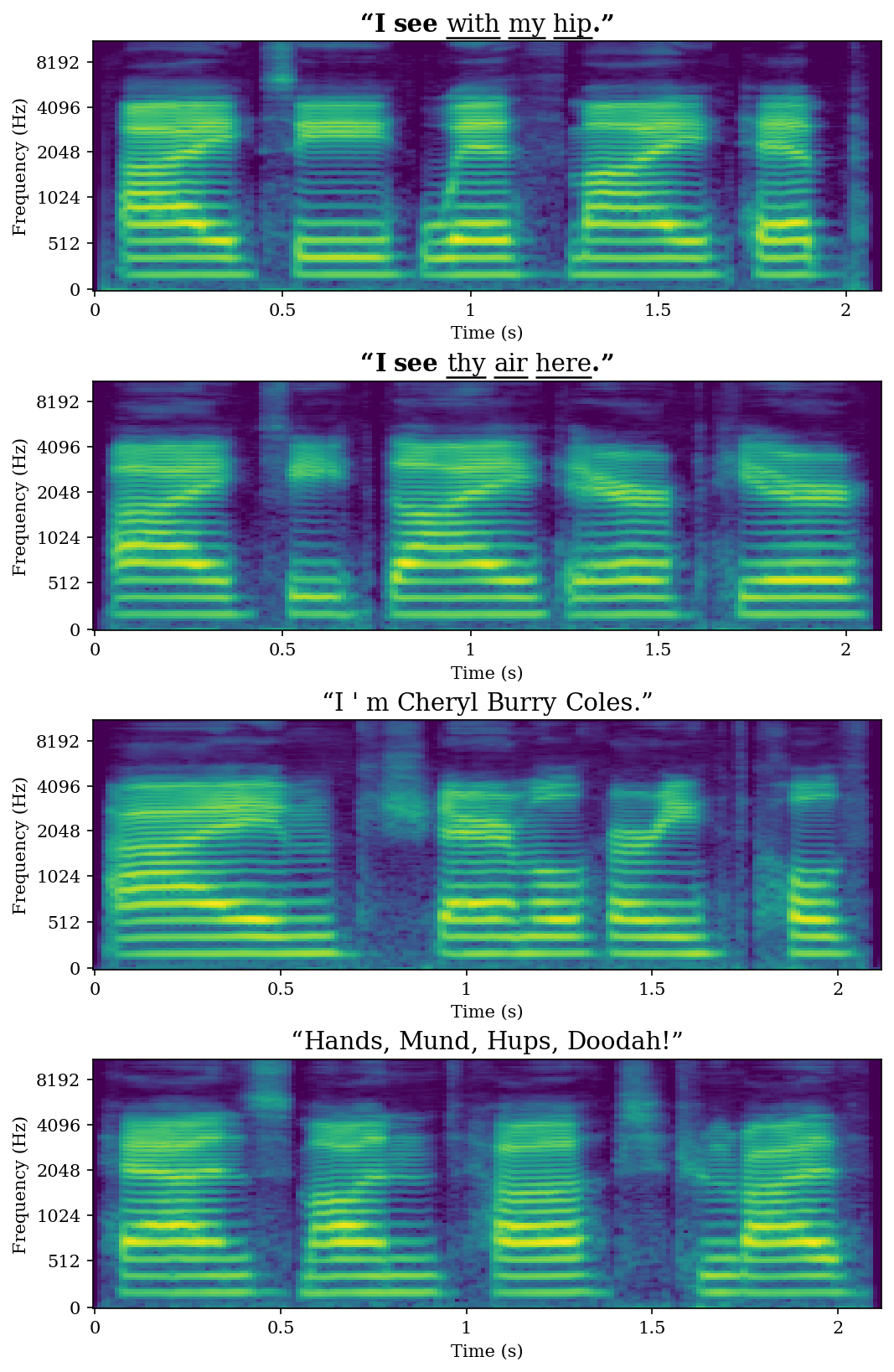}
    \end{minipage}
    
    \caption{Additional Melspectrogram Comparison}
    \label{fig:spectrogram_supp_3}
\end{figure}

\begin{figure}[t]
    \centering
    \begin{minipage}[c]{0.05\linewidth}
        \centering
        \vspace{0.0cm}
        \rotatebox{90}{\makebox[2.4cm][c]{\small GT}} \\[0pt]
        \rotatebox{90}{\makebox[2.4cm][c]{\small \textbf{SENSE}}} \\[0pt]
        \rotatebox{90}{\makebox[2.4cm][c]{\small FESDE}} \\[0pt]
        \rotatebox{90}{\makebox[2.4cm][c]{\small FE\_Phoneme}}
    \end{minipage}%
    \begin{minipage}[c]{0.45\linewidth}
        \includegraphics[width=\linewidth]{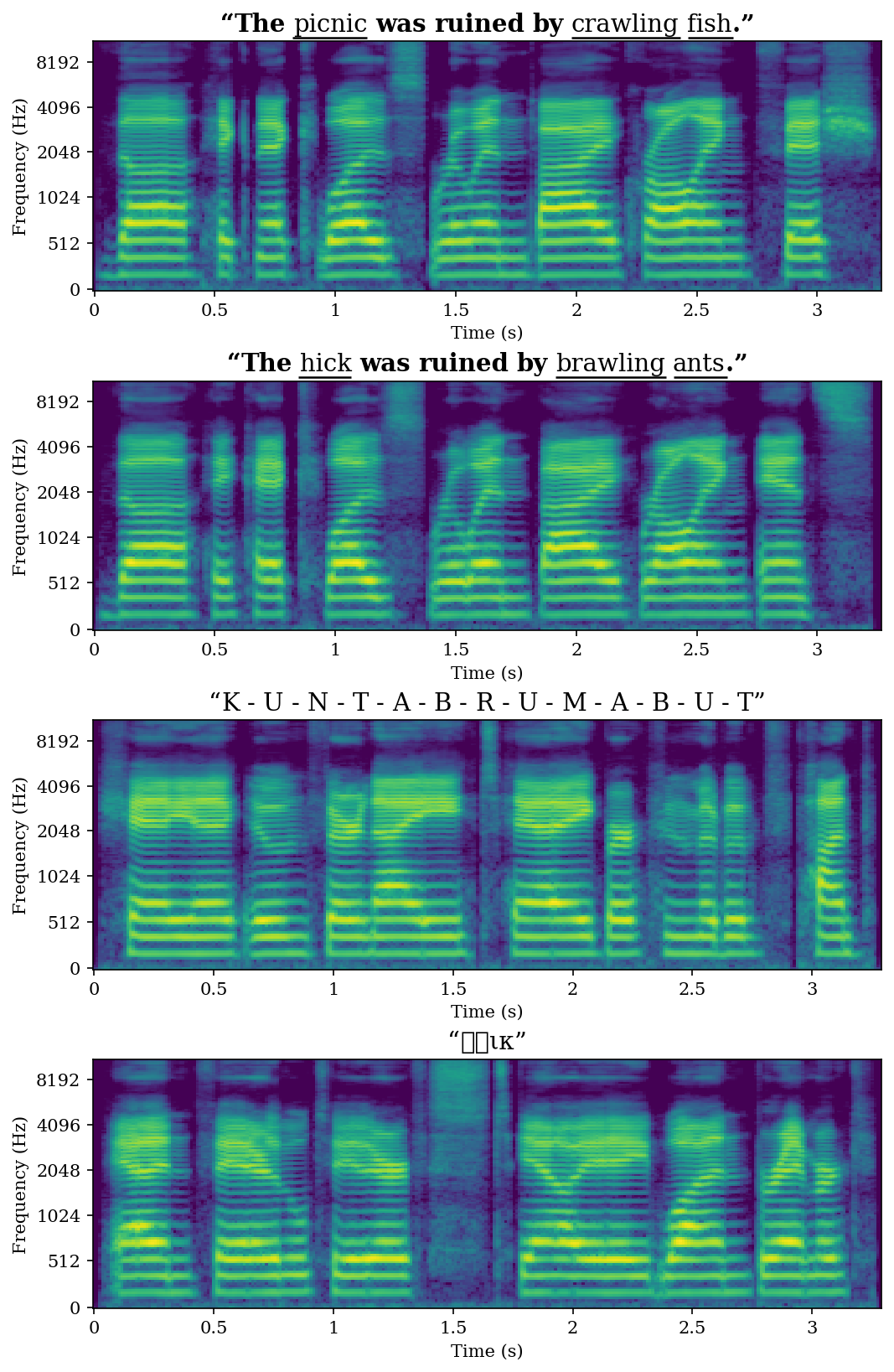}
    \end{minipage}
    \begin{minipage}[c]{0.45\linewidth}
        \includegraphics[width=\linewidth]{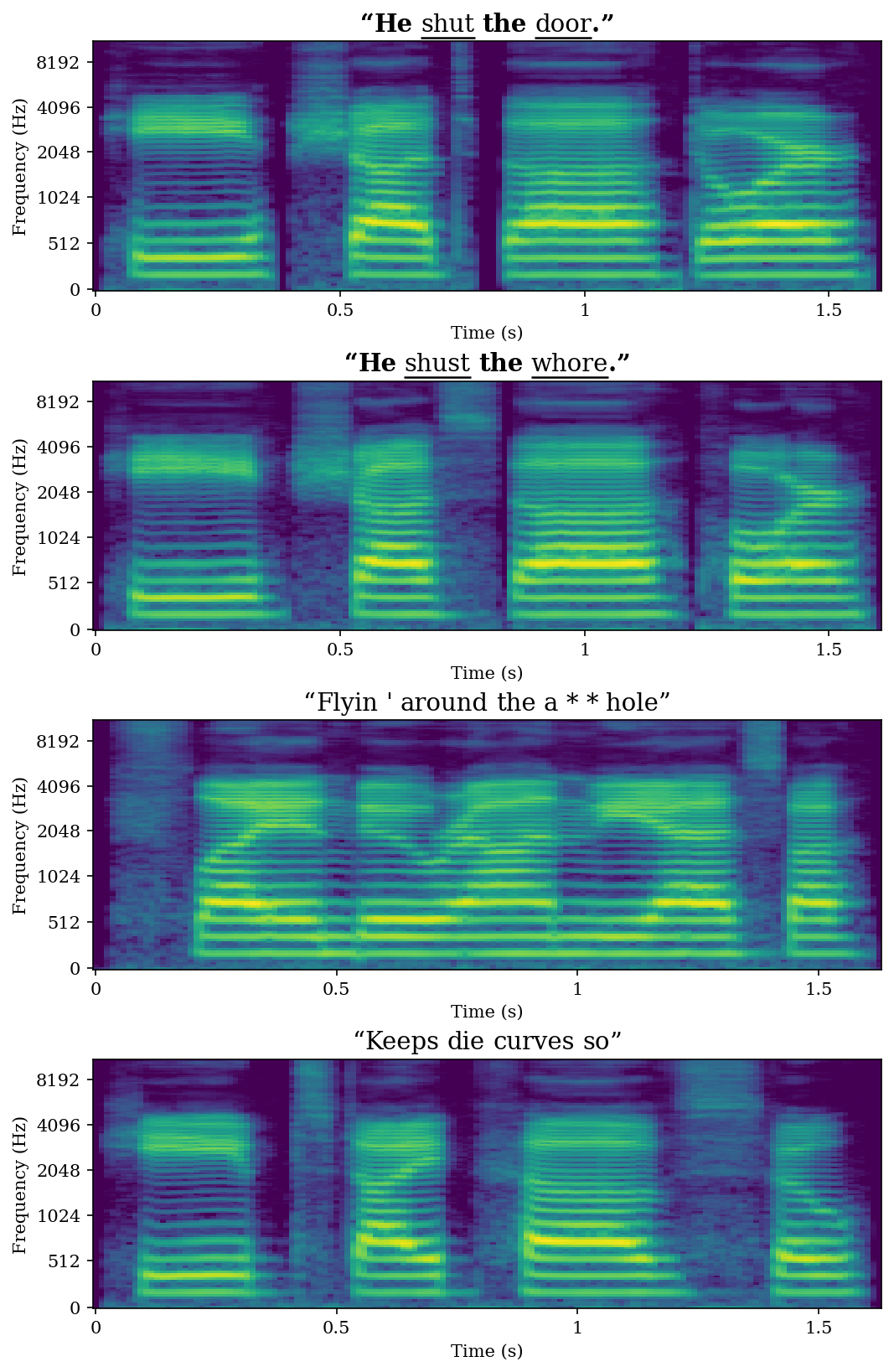}
    \end{minipage}

    \begin{minipage}[c]{0.05\linewidth}
        \centering
        \rotatebox{90}{\makebox[2.4cm][c]{\small GT}} \\[0pt]
        \rotatebox{90}{\makebox[2.4cm][c]{\small \textbf{SENSE}}} \\[0pt]
        \rotatebox{90}{\makebox[2.4cm][c]{\small FESDE}} \\[0pt]
        \rotatebox{90}{\makebox[2.4cm][c]{\small FE\_Phoneme}}
    \end{minipage}%
    \begin{minipage}[c]{0.45\linewidth}
        \includegraphics[width=\linewidth]{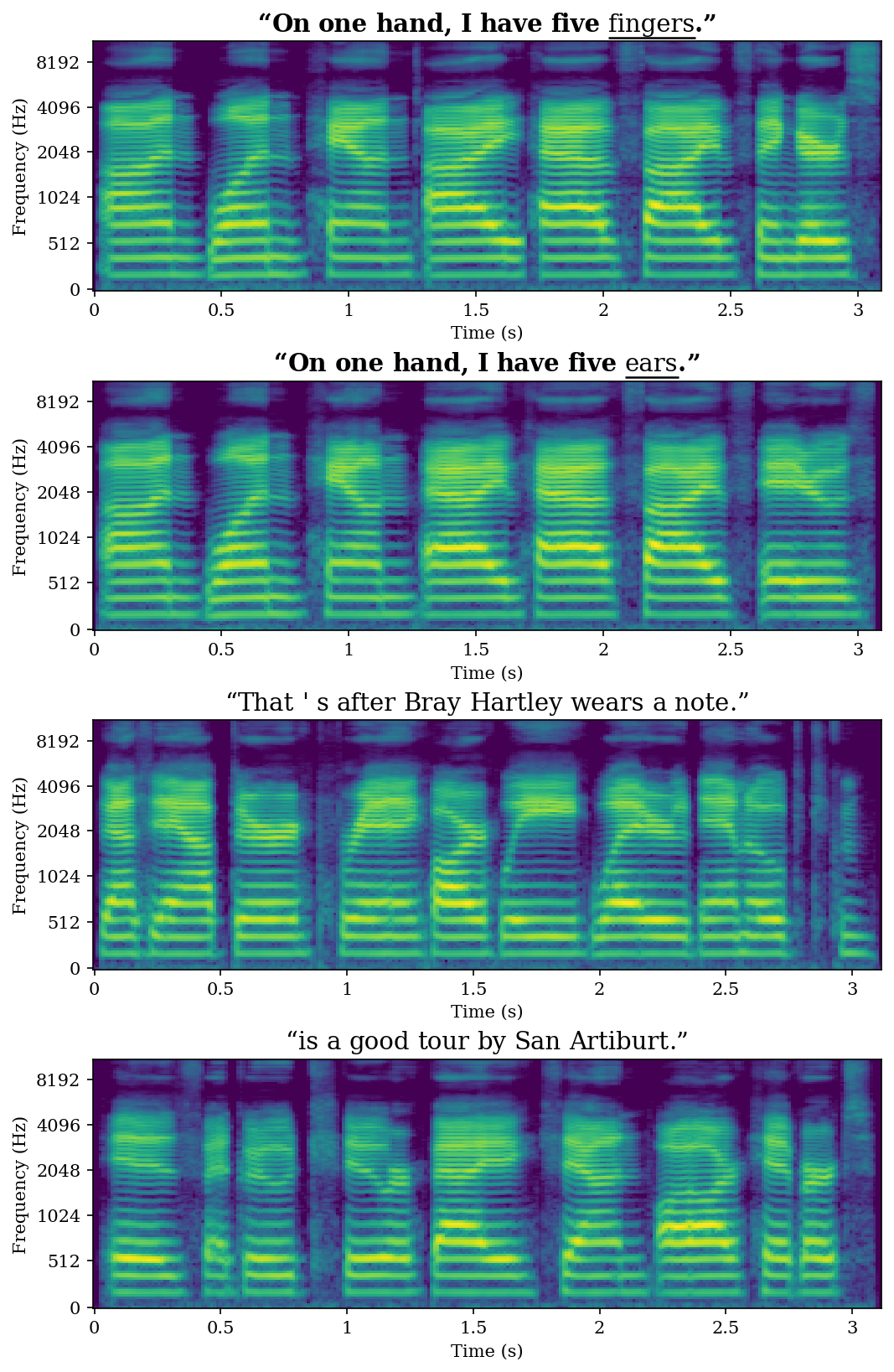}
    \end{minipage}%
    \begin{minipage}[c]{0.45\linewidth}
        \includegraphics[width=\linewidth]{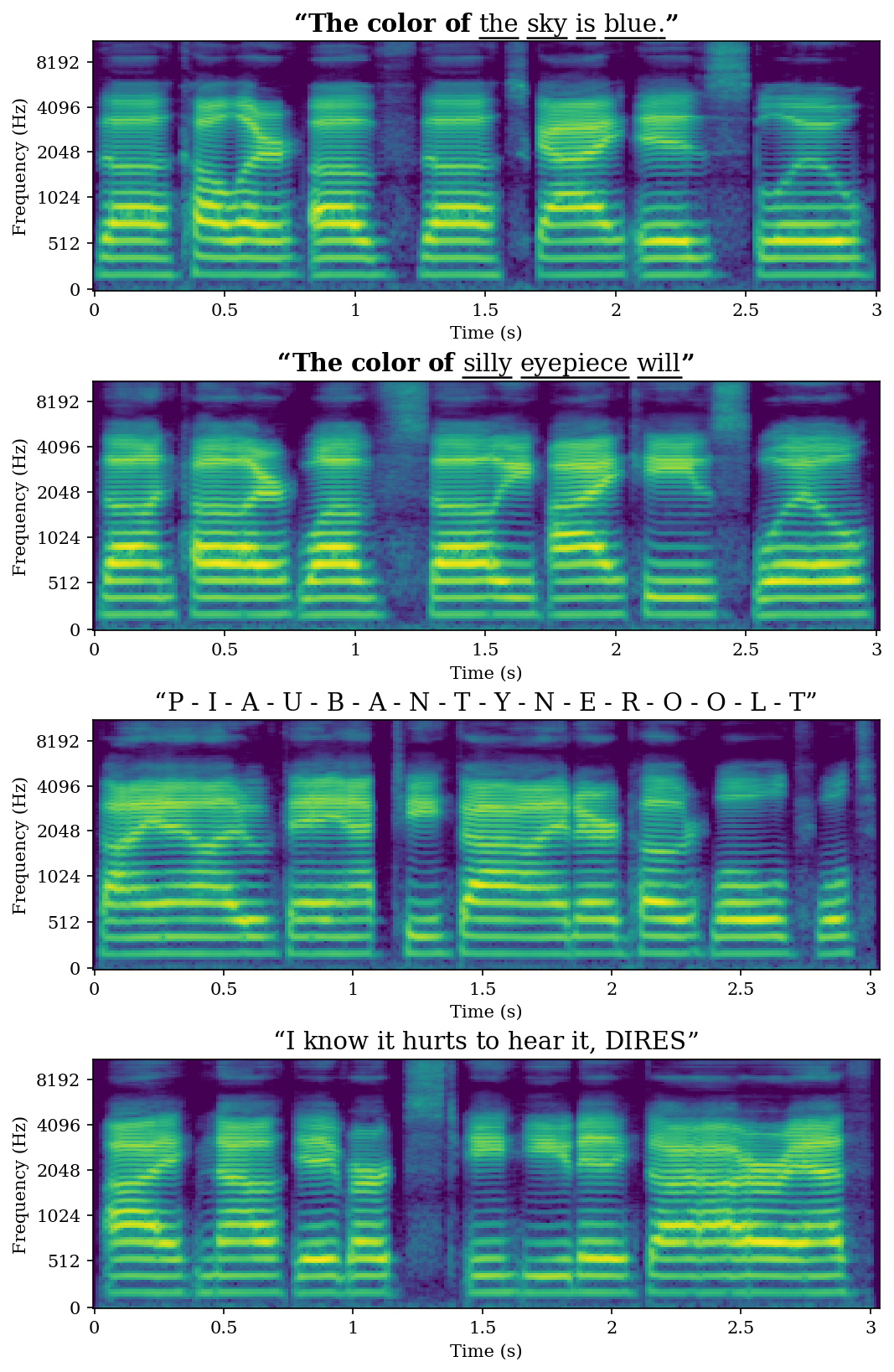}
    \end{minipage}
    
    \caption{Additional Melspectrogram Comparison}
    \label{fig:spectrogram_supp_4}
\end{figure}

\begin{figure}[t]
    \centering
    \begin{minipage}[c]{0.05\linewidth}
        \centering
        \vspace{0.0cm}
        \rotatebox{90}{\makebox[2.4cm][c]{\small GT}} \\[0pt]
        \rotatebox{90}{\makebox[2.4cm][c]{\small \textbf{SENSE}}} \\[0pt]
        \rotatebox{90}{\makebox[2.4cm][c]{\small FESDE}} \\[0pt]
        \rotatebox{90}{\makebox[2.4cm][c]{\small FE\_Phoneme}}
    \end{minipage}%
    \begin{minipage}[c]{0.45\linewidth}
        \includegraphics[width=\linewidth]{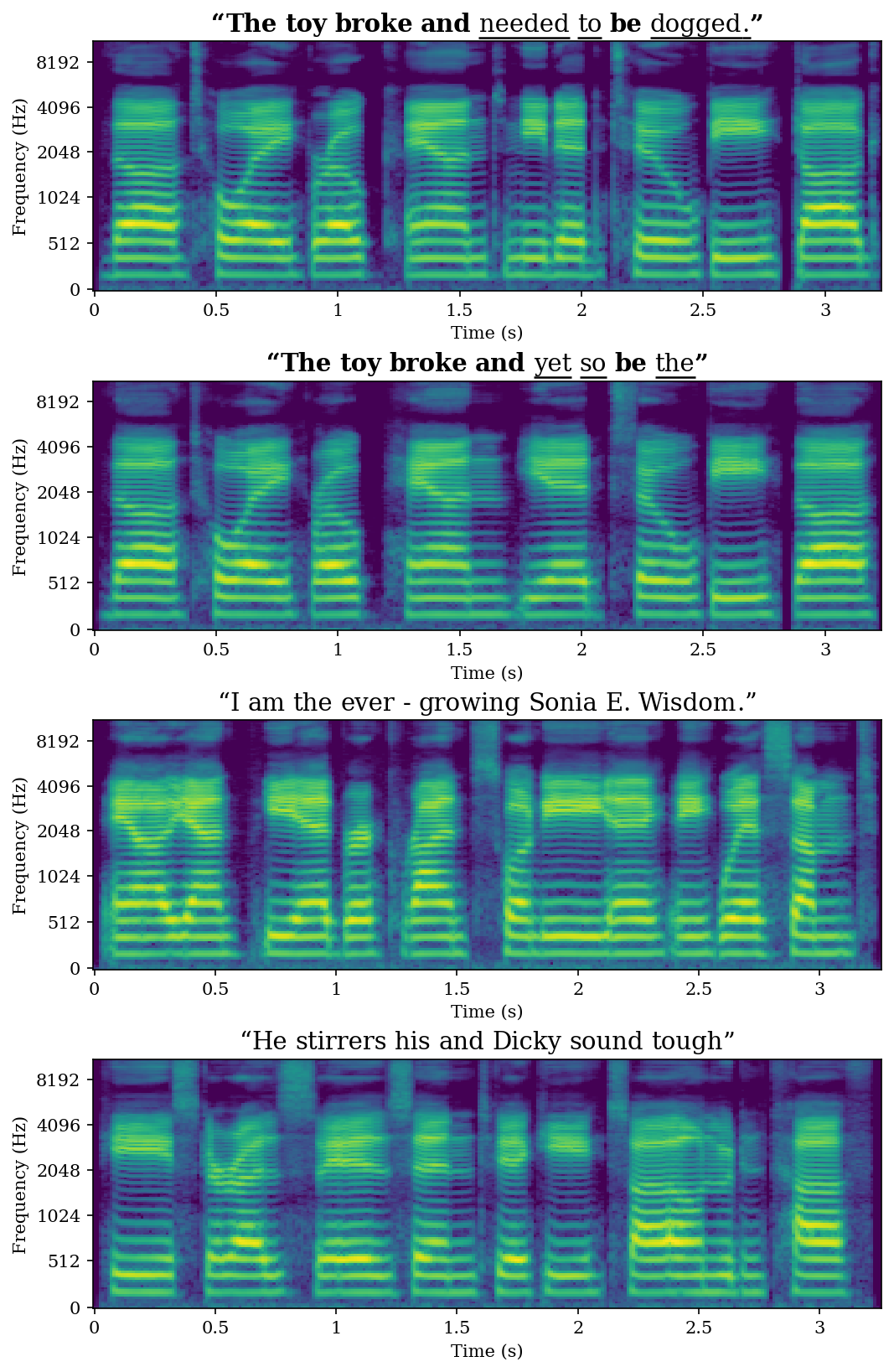}
    \end{minipage}
    \begin{minipage}[c]{0.45\linewidth}
        \includegraphics[width=\linewidth]{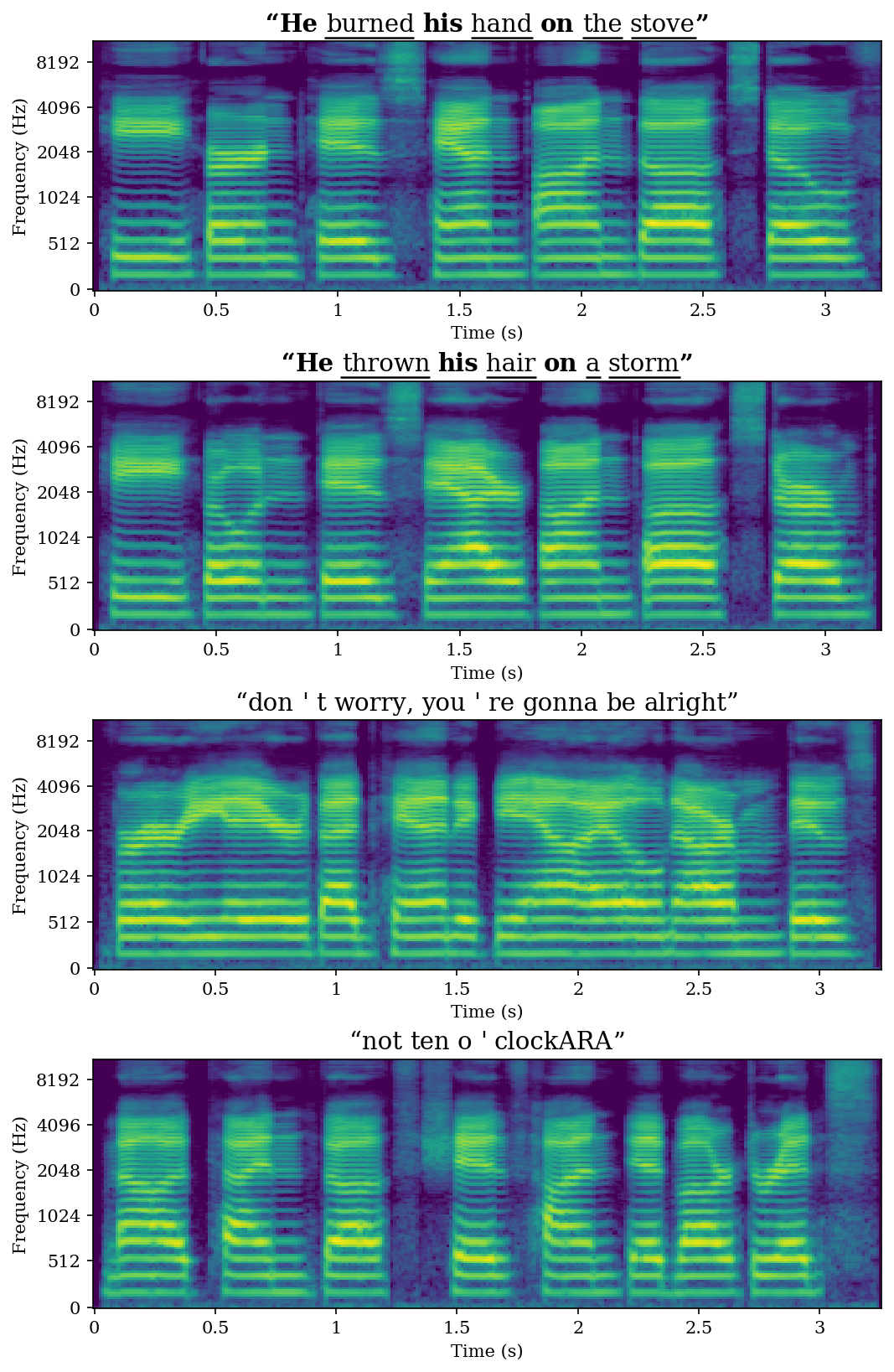}
    \end{minipage}

    \begin{minipage}[c]{0.05\linewidth}
        \centering
        \rotatebox{90}{\makebox[2.4cm][c]{\small GT}} \\[0pt]
        \rotatebox{90}{\makebox[2.4cm][c]{\small \textbf{SENSE}}} \\[0pt]
        \rotatebox{90}{\makebox[2.4cm][c]{\small FESDE}} \\[0pt]
        \rotatebox{90}{\makebox[2.4cm][c]{\small FE\_Phoneme}}
    \end{minipage}%
    \begin{minipage}[c]{0.45\linewidth}
        \includegraphics[width=\linewidth]{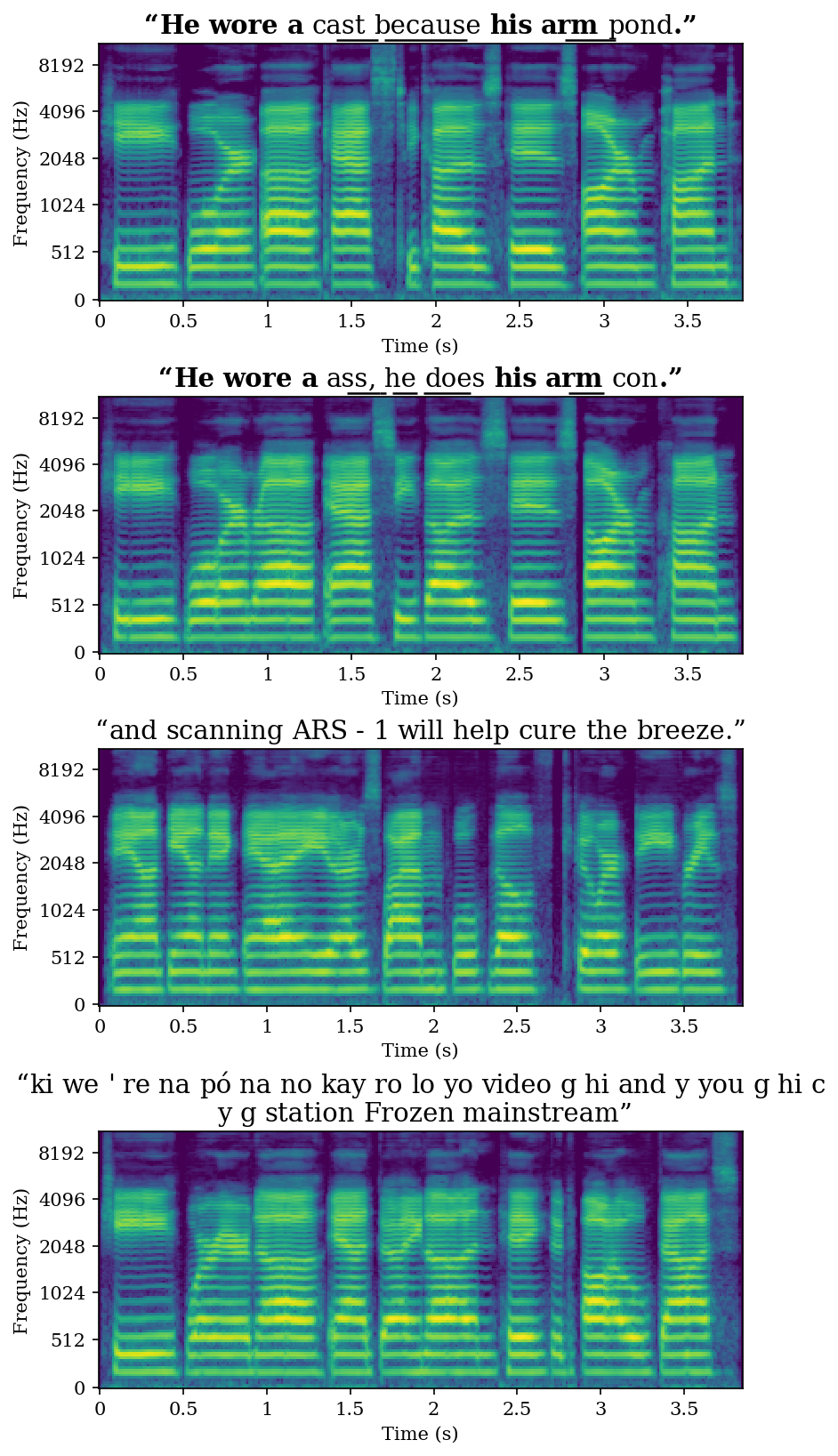}
    \end{minipage}%
    
    \caption{Additional Melspectrogram Comparison}
    \label{fig:spectrogram_supp_5}
\end{figure}

We provide more visual comparison samples in Figures~\ref{fig:spectrogram_supp_1},~\ref{fig:spectrogram_supp_2},~\ref{fig:spectrogram_supp_3},~\ref{fig:spectrogram_supp_4}, and~\ref{fig:spectrogram_supp_5}.
%------------------------------------------------------------------

\subsection{Validating the Semantic Role of ESC}
\label{app:semantic_validation}

To further investigate whether EEG Semantic Conditioning (ESC) captures sentence-specific information rather than merely improving acoustic reconstruction, we conduct additional analyses at both the representation and generation levels. Specifically, we examine (i) trial-specific EEG--text correspondence, (ii) sentence retrieval, (iii) the effect of matched and mismatched semantic conditioning, and (iv) the training-time and inference-time contributions of ESC.

\subsubsection{Trial-Specific EEG--Text Correspondence}

We first examine whether the EEG-derived representation $g_{\mathrm{eeg}}$ is more closely aligned with its corresponding sentence embedding than with unrelated sentences. For each test trial, we compute cosine similarity between $g_{\mathrm{eeg}}$ and the matched CLIP text embedding $g_{\mathrm{clip}}$, and compare it with the mean similarity to the remaining 439 sentence embeddings among 440 candidates.

\begin{table}[h]
\centering
\caption{EEG--text correspondence on held-out trials. The gap denotes the difference between matched and mismatched cosine similarities.}
\label{tab:semantic_alignment}
\small
\begin{tabular}{lccc}
\toprule
Evaluation subset & Matched & Mismatched & Gap \\
\midrule
All trials ($N=800$) & 0.37 & 0.25 & +0.12 \\
Congruent ($N=400$) & 0.40 & 0.25 & +0.15 \\
\bottomrule
\end{tabular}
\end{table}

As shown in Table~\ref{tab:semantic_alignment}, the matched sentence consistently exhibits higher similarity than mismatched sentences. The difference is statistically significant for both all trials ($p<10^{-79}$) and congruent trials ($p<10^{-45}$), with a larger gap for the congruent trials used for ESC supervision. These results indicate that the learned EEG representation contains trial-specific information aligned with the corresponding text embeddings.

\subsubsection{Sentence Retrieval}

We further evaluate whether $g_{\mathrm{eeg}}$ can identify the corresponding sentence among multiple candidates. Using 400 held-out congruent trials as queries, we rank 440 unique sentence embeddings by cosine similarity to the EEG-derived representation.

\begin{table}[h]
\centering
\caption{EEG-to-text retrieval among 440 candidate sentences. Random chance corresponds to uniformly random ranking.}
\label{tab:semantic_retrieval}
\small
\begin{tabular}{lccc}
\toprule
Method & Top-1 & Top-10 & Top-20 \\
\midrule
EEG retrieval & 2.8\% & 15.5\% & 23.0\% \\
Random chance & 0.23\% & 2.27\% & 4.55\% \\
\bottomrule
\end{tabular}
\end{table}

The EEG-derived representation achieves above-chance retrieval at all evaluated cutoffs. Although absolute retrieval accuracy remains modest, these results provide additional evidence of sentence-specific EEG--text correspondence rather than alignment solely to a shared distribution of text embeddings.

\subsubsection{Matched and Mismatched Semantic Conditioning}

To investigate whether the learned semantic representation contributes to speech generation, we intervene on the conditioning input at inference time while keeping all other inputs fixed. We compare the original EEG-derived conditioning $g_{\mathrm{eeg}}$ with the matched CLIP embedding and a mismatched CLIP embedding from another trial. Mismatched embeddings are assigned using a deterministic cyclic shift without self-pairs.

\begin{table}[h]
\centering
\caption{Effect of semantic conditioning on speech generation. Lower MCD and higher Mel-Corr and BERT are better.}
\label{tab:semantic_intervention}
\small
\begin{tabular}{lccc}
\toprule
Conditioning & MCD $\downarrow$ & Mel-Corr $\uparrow$ & BERT $\uparrow$ \\
\midrule
EEG-derived & 9.942 & 32.14 & 0.096 \\
Matched CLIP & 9.778 & 32.81 & 0.097 \\
Mismatched CLIP & 9.919 & 32.14 & 0.092 \\
\bottomrule
\end{tabular}
\end{table}

As shown in Table~\ref{tab:semantic_intervention}, matched conditioning produces the highest BERTScore, followed by EEG-derived and mismatched conditioning. However, the differences are small, and the acoustic metrics do not exhibit a consistent separation between EEG-derived and mismatched conditioning. These findings suggest that conditioning identity has a modest effect on sentence-level semantic similarity. They do not establish reliable recovery of sentence meaning in the generated speech.

\subsubsection{Training-Time and Inference-Time Effects}

Finally, we distinguish the effect of supplying semantic conditioning at inference from the representation-shaping effect of ESC during training. For the inference-time analysis, we replace $g_{\mathrm{eeg}}$ with the learned null embedding in the ESC-trained model. For the training-time analysis, we compare models trained with and without ESC under identical null conditioning.

\begin{table}[h]
\centering
\caption{Contributions of ESC at inference and during training. Positive values indicate improvements from EEG-derived conditioning or ESC training, respectively.}
\label{tab:esc_effects}
\small
\begin{tabular}{lcc}
\toprule
Comparison & $\Delta$ Mel-Corr & $\Delta$ BERTScore \\
\midrule
EEG-derived vs.\ null & +0.11 & +0.002 \\
ESC vs.\ no ESC (null) & +2.38 & +0.004 \\
\bottomrule
\end{tabular}
\end{table}

Replacing EEG-derived conditioning with the null embedding results in a small decrease in semantic similarity, indicating a modest inference-time effect. Importantly, the ESC-trained model retains an advantage over the model trained without ESC even when both receive null conditioning. Since the ESC objective backpropagates through the semantic projection and EEG encoder, this result suggests that ESC also shapes the learned EEG representation during training.

Together, these experiments indicate two complementary roles of ESC: representation shaping during training and modest semantic guidance during inference. While the learned representation exhibits trial-specific EEG--text correspondence, the current reconstruction quality does not support reliable semantic speech decoding.

\subsection{Generalization Beyond the N400 Paradigm}
\label{app:natural_speech}

The N400 dataset uses artificially constructed sentences with controlled timing and limited prosodic variation. Although this setting facilitates EEG--stimulus alignment and comparison with prior work, it differs substantially from continuous natural speech. We therefore investigate whether ESC requires congruency-based supervision and whether the learned EEG representation can be adapted to naturalistic listened speech.

\subsubsection{ESC Without Congruency-Based Selection}

While the main SENSE model applies ESC only to congruent trials, the ESC objective itself requires paired EEG--text samples rather than explicit congruency labels. To evaluate the role of this restriction, we additionally train SENSE by applying ESC to all training trials, including both congruent and incongruent sentences.

\begin{table}[h]
\centering
\caption{Effect of ESC supervision selection.}
\label{tab:esc_supervision}
\small
\begin{tabular}{lccccc}
\toprule
ESC supervision & MCD $\downarrow$ & Mel-Corr $\uparrow$ & STOI $\uparrow$ & WER $\downarrow$ & BERT $\uparrow$ \\
\midrule
Congruent only & 9.93 & 32.07 & 0.413 & 0.995 & 0.096 \\
All trials & 10.17 & 30.90 & 0.409 & 1.018 & 0.094 \\
Without ESC & 10.34 & 29.43 & 0.401 & 1.032 & 0.090 \\
\bottomrule
\end{tabular}

\end{table}

Applying ESC to all trials improves all five metrics compared with removing ESC entirely. Restricting supervision to congruent trials provides further improvements, suggesting that congruency serves as a dataset-specific supervision filter rather than an architectural requirement.

\subsubsection{Transfer to Continuous Natural Speech}

To further examine generalization beyond controlled N400 stimuli, we evaluate SENSE on the Broderick naturalistic speech dataset~\cite{broderick2018electrophysiological}, in which participants listened to continuous narrative speech. We initialize the EEG encoder and ESC projection using the N400-trained weights and fine-tune them on the naturalistic dataset. Each EEG segment is paired with its corresponding text representation, and we evaluate the cosine similarity between matched EEG--text pairs and randomly mismatched pairs.

Because the public release available to us does not provide the aligned speech waveforms required for reconstruction, we evaluate representation-level EEG--text alignment rather than speech synthesis.

\begin{table}[h]
\centering
\caption{EEG--text alignment on the Broderick naturalistic speech dataset.}
\label{tab:natural_speech}
\small
\begin{tabular}{lccc}
\toprule
Dataset & Matched & Random & Gap \\
\midrule
Broderick & 0.602 & 0.517 & +0.085 \\
\bottomrule
\end{tabular}
\end{table}

Matched EEG--text similarity exceeds randomly mismatched similarity for all 19 participants. Furthermore, none of 5,000 random permutations produces a similarity gap as large as the observed difference ($p<2.0\times10^{-4}$). These results demonstrate that the N400-pretrained encoder and ESC objective can be adapted to continuous listened-speech EEG without congruency labels or artificially segmented sentence stimuli.

Nevertheless, this experiment evaluates transfer at the representation level and does not establish naturalistic speech reconstruction performance. Full EEG-to-speech evaluation on naturalistic datasets with aligned speech waveforms remains an important direction for future work.

\subsection{Channel Attention Gate Visualization}
\label{app:gate_viz}

\begin{figure}[t]
    \centering
    \includegraphics[width=1.0\linewidth]{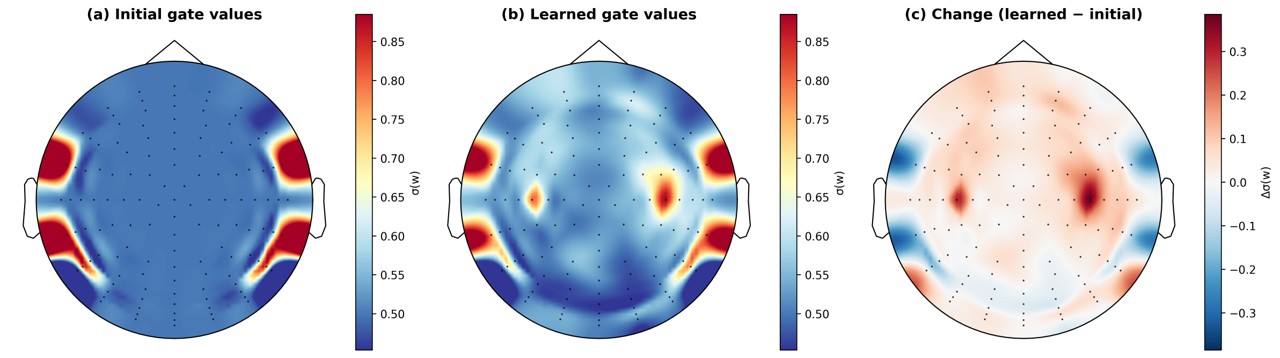}
    \caption{\textbf{Channel attention gate before and after training.} Sigmoid-gated weights $\sigma(\mathbf{w}) \in (0, 1)$ projected onto the BioSemi 128-channel scalp layout. (a) Initialization: positive bias ($\sigma(\mathbf{w}) \approx 0.88$) on six auditory-related electrodes (T7, T8, FT7, FT8, TP7, TP8); all other channels start neutral ($\sigma(\mathbf{w}) = 0.5$). (b) Values after training. (c) Per-channel change $\Delta\sigma(\mathbf{w})$. Auditory-prior electrodes remain among the most activated but are modestly attenuated. The largest gains appear near central electrodes C3 and C4 ($\Delta\sigma \approx +0.27$ and $+0.29$), with additional gains in frontal-central regions; Occipital channels, which subserve visual processing and are not expected to contribute to auditory tasks, are suppressed below their initial value, indicating that the gate not only emphasizes auditory-related electrodes but also actively down-weights electrodes unrelated to the task.}
    \label{fig:gate}
\end{figure}

Figure~\ref{fig:gate} shows the channel attention gate values $\sigma(\mathbf{w})$ projected onto the BioSemi 128-channel scalp layout, comparing the initialization (positive bias on six auditory-related electrodes, zero elsewhere) with the values after training and the per-channel change.

%------------------------------------------------------------------
\subsection{Attribution Trajectory During Training}
\label{app:attribution_trajectory}

\begin{figure}[t]
    \centering
    \includegraphics[width=1.0\linewidth]{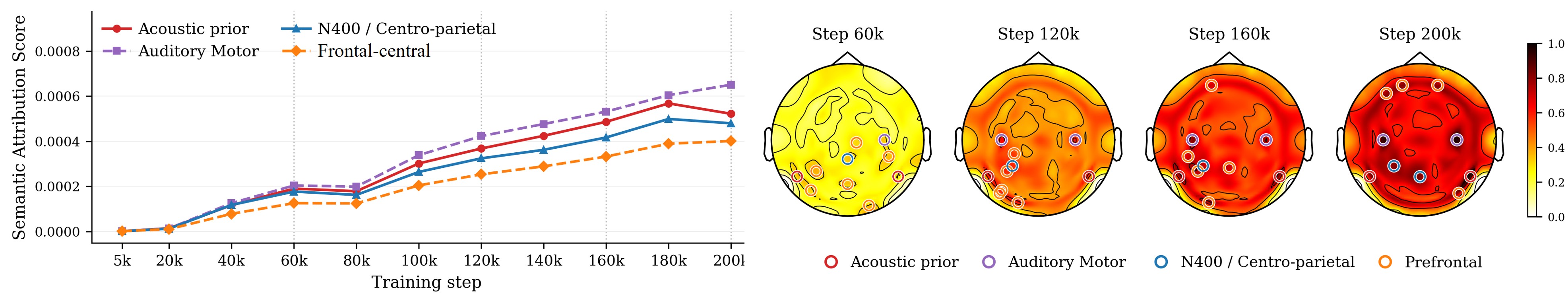}
    \caption{\textbf{Semantic channel importance trajectory during training.} (Left) Gradient-based attribution score per EEG ROI group across training steps. Vertical dotted lines indicate steps shown in the topomaps. (Right) Spatial distribution of semantic attribution at selected checkpoints; brighter regions indicate higher channel importance for EEG–text alignment. The model initially relies on acoustic prior regions, and progressively shifts attribution toward N400/centro-parietal channels associated with semantic processing.}
    \label{fig:topomap}
\end{figure}

Figure~\ref{fig:topomap} tracks the gradient-based attribution across training. Early in training, attribution is concentrated over auditory-prior regions reflecting the gate initialization. As training progresses, attribution to centro-parietal and frontal-central regions increases, while auditory contributions remain present but no longer dominant.

% ===== TOPO

\subsection{Sensitivity to Loss Weights}
\label{app:lambda_sensitivity}
We examine sensitivity to the ESC and CTC loss weights ($\lambda_{\mathrm{esc}}$ and $\lambda_{\mathrm{ctc}}$). Table~\ref{tab:lambda_sensitivity} shows the chosen values balance the reconstruction quality and semantic quality, yielding best results.

\begin{table}[t]
\centering
\caption{\textbf{Sensitivity analysis of loss weights $\lambda_\text{ctc}$ and $\lambda_\text{esc}$.}
Results are reported on the \textit{subject} split (mean $\pm$ std over 5 runs).}
\label{tab:lambda_sensitivity}
\small
\setlength{\tabcolsep}{10pt}
\begin{tabular}{cc ccccc}
\toprule
$\lambda_\text{ctc}$ & $\lambda_\text{esc}$
  & MCD $\downarrow$ & Mel-Corr $\uparrow$ & STOI $\uparrow$ & WER $\downarrow$ & CLIP-Sim $\uparrow$ \\
\midrule
0.0 & 0.5
  & 10.50\tiny{$\pm$0.02} & 27.24\tiny{$\pm$0.28} & 0.372 & 1.050& 0.756 \\
0.1 & 0.5
  & 10.27\tiny{$\pm$0.03} & 27.92\tiny{$\pm$0.24} & 0.384 & 1.041 & 0.755 \\
\textbf{0.3} & \textbf{0.5}
  & \textbf{9.93}\tiny{$\pm$0.03} & \textbf{32.07}\tiny{$\pm$0.15} & \textbf{0.413} & \textbf{0.994} & \textbf{0.760} \\
1.0 & 0.5
  & 10.01\tiny{$\pm$0.03} & 30.39\tiny{$\pm$0.15} & 0.404 & 1.002 & 0.755 \\
\midrule
0.3 & 0.0
  & 10.34\,{\tiny$\pm$0.03} & 29.43\,{\tiny$\pm$0.27} & 0.402 & 1.032 & 0.760\\
0.3 & 0.1
  & 10.33\,{\tiny$\pm$0.01} & 29.03\,{\tiny$\pm$0.24} & 0.392 & 1.064 & 0.759 \\
\textbf{0.3} & \textbf{0.5}
  & \textbf{9.93}\tiny{$\pm$0.03} & \textbf{32.07}\tiny{$\pm$0.15} & \textbf{0.413} & \textbf{0.994}& \textbf{0.760}\\
0.3 & 2.0
  & 10.07\,{\tiny$\pm$.02} & 29.86\,{\tiny$\pm$0.24} & 0.402 & 1.030 & 0.757\\
\bottomrule
\end{tabular}
\end{table}

\subsection{Per-Subject Results}
\label{app:per_subject}

Table~\ref{tab:per_subject_full} reports the complete per-subject results for the cross-subject generalization experiments shown in Figure~\ref{fig:cross_subject}. For each of the three held-out pairs (sub-1, sub-2), (sub-7, sub-8), and (sub-23, sub-24), all models are retrained from scratch on the remaining 18 training subjects and evaluated on each held-out subject individually. All values are mean $\pm$ standard deviation over 5 random seeds. SENSE outperforms both baselines on every metric and every held-out subject.

\begin{table*}[t]
\centering
\caption{\textbf{Complete per-subject results for cross-subject generalization.}
For each held-out pair, all models are retrained from scratch on the remaining 18 subjects.
Bold: best per subject.}
\label{tab:per_subject_full}
\setlength{\tabcolsep}{13pt}
\scriptsize{
\begin{tabular}{llccccc}
\toprule
\textbf{Held-out}
& \textbf{Model}
& \textbf{MCD}$\downarrow$
& \textbf{Mel-Corr}$\uparrow$
& \textbf{STOI}$\uparrow$
& \textbf{WER}$\downarrow$
& \textbf{BERT}$\uparrow$ \\
\midrule

\multirow{3}{*}{sub-1}
& FESDE
& 11.847\,{\tiny$\pm$0.04}
& 19.282\,{\tiny$\pm$0.39}
& 0.2793
& 1.1851
& 0.0150\\

& FE-Phoneme
& 11.095\,{\tiny$\pm$0.07}
& 24.920\,{\tiny$\pm$0.51}
& 0.3518
& 1.1694
& 0.0318 \\

& \textbf{SENSE}
& \textbf{10.226}\,{\tiny$\pm$0.01}
& \textbf{29.671}\,{\tiny$\pm$0.36}
& \textbf{0.4008}
& \textbf{1.0289}
& \textbf{0.0902} \\
\midrule

\multirow{3}{*}{sub-2}
& FESDE
& 11.920\,{\tiny$\pm$0.08}
& 19.407\,{\tiny$\pm$0.57}
& 0.2780
& 1.1939
& 0.0256 \\

& FE-Phoneme
& 11.170\,{\tiny$\pm$0.08}
& 24.352\,{\tiny$\pm$0.47}
& 0.3442
& 1.1508
& 0.0283 \\

& \textbf{SENSE}
& \textbf{10.175}\,{\tiny$\pm$0.03}
& \textbf{30.311}\,{\tiny$\pm$0.20}
& \textbf{0.4065}
& \textbf{1.0088}
& \textbf{0.0965}\\
\midrule

\multirow{3}{*}{sub-7}
& FESDE
& 11.659\,{\tiny$\pm$0.05}
& 19.836\,{\tiny$\pm$0.52}
& 0.2765
& 1.1998
& 0.0371 \\

& FE-Phoneme
& 11.127\,{\tiny$\pm$0.05}
& 25.029\,{\tiny$\pm$0.20}
& 0.3529
& 1.1731
& 0.0318 \\

& \textbf{SENSE}
& \textbf{10.156}\,{\tiny$\pm$0.03}
& \textbf{29.392}\,{\tiny$\pm$0.17}
& \textbf{0.4009}
& \textbf{1.0070}
& \textbf{0.0909}\\
\midrule

\multirow{3}{*}{sub-8}
& FESDE
& 11.609\,{\tiny$\pm$0.08}
& 19.438\,{\tiny$\pm$0.43}
& 0.2726
& 1.2279
& 0.0419\\

& FE-Phoneme
& 11.134\,{\tiny$\pm$0.06}
& 24.858\,{\tiny$\pm$0.45}
& 0.3583
& 1.1640
& 0.0194 \\

& \textbf{SENSE}
& \textbf{10.159}\,{\tiny$\pm$0.02}
& \textbf{30.005}\,{\tiny$\pm$0.30}
& \textbf{0.4041}
& \textbf{1.0040}
& \textbf{0.0826} \\
\midrule

\multirow{3}{*}{sub-23}
& FESDE
& 11.726\,{\tiny$\pm$0.10}
& 18.978\,{\tiny$\pm$0.42}
& 0.2774
& 1.2106
& 0.0310 \\

& FE-Phoneme
& 10.645\,{\tiny$\pm$0.04}
& 27.118\,{\tiny$\pm$0.16}
& 0.3833
& 1.1815
& 0.0342 \\

& \textbf{SENSE}
& \textbf{10.037}\,{\tiny$\pm$0.03}
& \textbf{31.547}\,{\tiny$\pm$0.12}
& \textbf{0.4092}
& \textbf{1.0132}
& \textbf{0.0947}\\
\midrule

\multirow{3}{*}{sub-24}
& FESDE
& 11.736\,{\tiny$\pm$0.10}
& 19.103\,{\tiny$\pm$0.54}
& 0.2762
& 1.1977
& 0.0318\\

& FE-Phoneme
& 10.652\,{\tiny$\pm$0.06}
& 26.690\,{\tiny$\pm$0.25}
& 0.3777
& 1.1476
& 0.0419 \\

& \textbf{SENSE}
& \textbf{9.852}\,{\tiny$\pm$0.02}
& \textbf{32.557}\,{\tiny$\pm$0.29}
& \textbf{0.4168}
& \textbf{0.9903}
& \textbf{0.1077}\\
\bottomrule
\end{tabular}
}
\end{table*}

%%%%%%%%%%%%%%%%%%%%%%%%%%%%%%%%%%%%%%%%%%%%%%%%%%%%%%%%%%%%

\clearpage
\section*{NeurIPS Paper Checklist}

%%% BEGIN INSTRUCTIONS %%%
%%% END INSTRUCTIONS %%%

\begin{enumerate}

\item {\bf Claims}
    \item[] Question: Do the main claims made in the abstract and introduction accurately reflect the paper's contributions and scope?
    \item[] Answer: \answerYes{} % Replace by \answerYes{}, \answerNo{}, or \answerNA{}.
    \item[] Justification: The abstract and introduction clearly state the paper's contributions: (1) reformulating EEG-to-speech as a joint acoustic-semantic problem, (2) proposing SENSE with a graph-based EEG encoder and EEG Semantic Conditioning (ESC), and (3) demonstrating improved acoustic and semantic metrics including strong cross-subject generalization. These claims are supported by experimental results in Section~\ref{sec:experiments}.

\item {\bf Limitations}
    \item[] Question: Does the paper discuss the limitations of the work performed by the authors?
    \item[] Answer: \answerYes{} % Replace by \answerYes{}, \answerNo{}, or \answerNA{}.
    \item[] Justification: Section~\ref{sec:conclusions} discusses that the N400 corpus uses fixed templates with varied sentence-final words, which differs from naturalistic speech, and that absolute WER values indicate practical EEG-to-speech reconstruction remains challenging. Extension to naturalistic speech and active-speech paradigms is a natural next step.

\item {\bf Theory assumptions and proofs}
    \item[] Question: For each theoretical result, does the paper provide the full set of assumptions and a complete (and correct) proof?
    \item[] Answer: \answerNA{} % Replace by \answerYes{}, \answerNo{}, or \answerNA{}.
    \item[] Justification: The paper does not include theoretical results requiring formal proofs.

    \item {\bf Experimental result reproducibility}
    \item[] Question: Does the paper fully disclose all the information needed to reproduce the main experimental results of the paper to the extent that it affects the main claims and/or conclusions of the paper (regardless of whether the code and data are provided or not)?
    \item[] Answer: \answerYes{} % Replace by \answerYes{}, \answerNo{}, or \answerNA{}.
    \item[] Justification:  Section~\ref{sec:method} fully describes the model architecture, and Appendix~\ref{app:impl} provides comprehensive implementation details including dataset preprocessing, architecture specifications, training configuration, evaluation protocol, and loss weights. Subject splits and the nested cross-subject scaling structure are also specified. Anonymized code is provided as supplementary material for direct reproduction.

\item {\bf Open access to data and code}
    \item[] Question: Does the paper provide open access to the data and code, with sufficient instructions to faithfully reproduce the main experimental results, as described in supplemental material?
    \item[] Answer: \answerYes{} % Replace by \answerYes{}, \answerNo{}, or \answerNA{}.
    \item[] Justification: The N400 EEG corpus~\cite{toffolo2022} is publicly available and cited. We provide anonymized code as supplementary material (a single zip file under 100MB). The README provides instructions for environment setup, data preparation, training, and inference. Pretrained checkpoints are not included due to file size constraints but will be released with the de-anonymized version upon acceptance.

\item {\bf Experimental setting/details}
    \item[] Question: Does the paper specify all the training and test details (e.g., data splits, hyperparameters, how they were chosen, type of optimizer) necessary to understand the results?
    \item[] Answer: \answerYes{} % Replace by \answerYes{}, \answerNo{}, or \answerNA{}.
    \item[] Justification: Section~\ref{sec:experiments} specifies dataset splits, baselines, metrics, and training setup. Appendix~\ref{app:training} details optimizer, learning rate, loss weights, training duration, and hardware. Hyperparameter selection is justified via the sensitivity analysis in Appendix ~\ref{app:lambda_sensitivity}

\item {\bf Experiment statistical significance}
    \item[] Question: Does the paper report error bars suitably and correctly defined or other appropriate information about the statistical significance of the experiments?
    \item[] Answer: \answerYes{} % Replace by \answerYes{}, \answerNo{}, or \answerNA{}.
    \item[] Justification: All main results in Tables~\ref{tab:main_results},~\ref{tab:ablation},~\ref{tab:graph_ablation},~\ref{tab:esc_encoder},~\ref{tab:lambda_sensitivity},~\ref{tab:per_subject_full} and Figures~\ref{fig:scaling_curve}, ~\ref{fig:cross_subject} report mean ± standard deviation over 5 random seeds. The variability captured is across random seeds with fixed train/test splits, as noted in Section~\ref{sec:experiments} and Appendix~\ref{app:training}.

\item {\bf Experiments compute resources}
    \item[] Question: For each experiment, does the paper provide sufficient information on the computer resources (type of compute workers, memory, time of execution) needed to reproduce the experiments?
    \item[] Answer: \answerYes{} % Replace by \answerYes{}, \answerNo{}, or \answerNA{}.
    \item[] Justification: Appendix~\ref{app:cost} states that training a single SENSE model takes approximately 2 days on 2 NVIDIA RTX 3090 GPUs with FP32 precision and batch size 16 per GPU.
    
\item {\bf Code of ethics}
    \item[] Question: Does the research conducted in the paper conform, in every respect, with the NeurIPS Code of Ethics \url{https://neurips.cc/public/EthicsGuidelines}?
    \item[] Answer: \answerYes{} % Replace by \answerYes{}, \answerNo{}, or \answerNA{}.
    \item[] Justification: The research conforms to the NeurIPS Code of Ethics. The paper uses a publicly available EEG dataset~\cite{toffolo2022}, does not involve new human subject data collection, and preserves anonymity in submission.

\item {\bf Broader impacts}
    \item[] Question: Does the paper discuss both potential positive societal impacts and negative societal impacts of the work performed?
    \item[] Answer: \answerYes{} % Replace by \answerYes{}, \answerNo{}, or \answerNA{}.
    \item[] Justification: Broader impact is discussed.
    
\item {\bf Safeguards}
    \item[] Question: Does the paper describe safeguards that have been put in place for responsible release of data or models that have a high risk for misuse (e.g., pre-trained language models, image generators, or scraped datasets)?
    \item[] Answer: \answerNA{} % Replace by \answerYes{}, \answerNo{}, or \answerNA{}.
    \item[] Justification: This work does not release pretrained models or scraped datasets posing high misuse risk. Future released checkpoints (upon acceptance) will be trained on a publicly available EEG corpus and pose no high misuse risk, as discussed in Appendix~\ref{app:impact}.

\item {\bf Licenses for existing assets}
    \item[] Question: Are the creators or original owners of assets (e.g., code, data, models), used in the paper, properly credited and are the license and terms of use explicitly mentioned and properly respected?
    \item[] Answer: \answerYes{} % Replace by \answerYes{}, \answerNo{}, or \answerNA{}.
    \item[] Justification: All existing assets are properly cited with their original publications. Specifically, the N400 EEG corpus~\cite{toffolo2022} is released under CC BY-NC-ND 4.0 and is used for non-commercial research as permitted. The codebase builds upon publicly available implementations of VITS, S4, and Conformer, all of which are released under permissive open-source licenses (MIT or Apache 2.0). Our released code is provided under the MIT License, included as a LICENSE file in the supplementary material.

\item {\bf New assets}
    \item[] Question: Are new assets introduced in the paper well documented and is the documentation provided alongside the assets?
    \item[] Answer: \answerYes{} % Replace by \answerYes{}, \answerNo{}, or \answerNA{}.
    \item[] Justification:  We release anonymized code as supplementary material. The repository includes the model implementation, configuration files, training and inference scripts, a documented README, and a LICENSE file (MIT). Pretrained checkpoints are not included due to file size constraints but will be provided in the de-anonymized release upon acceptance.

\item {\bf Crowdsourcing and research with human subjects}
    \item[] Question: For crowdsourcing experiments and research with human subjects, does the paper include the full text of instructions given to participants and screenshots, if applicable, as well as details about compensation (if any)? 
    \item[] Answer: \answerNA{} % Replace by \answerYes{}, \answerNo{}, or \answerNA{}.
    \item[] Justification: The paper uses a publicly available EEG dataset~\cite{toffolo2022} and does not involve new crowdsourcing or human subject data collection.

\item {\bf Institutional review board (IRB) approvals or equivalent for research with human subjects}
    \item[] Question: Does the paper describe potential risks incurred by study participants, whether such risks were disclosed to the subjects, and whether Institutional Review Board (IRB) approvals (or an equivalent approval/review based on the requirements of your country or institution) were obtained?
    \item[] Answer: \answerNA{} % Replace by \answerYes{}, \answerNo{}, or \answerNA{}.
    \item[] Justification:  No new human subject research was conducted. The N400 EEG corpus~\cite{toffolo2022} is publicly available and was collected by the original authors under their institutional approval.

\item {\bf Declaration of LLM usage}
    \item[] Question: Does the paper describe the usage of LLMs if it is an important, original, or non-standard component of the core methods in this research? Note that if the LLM is used only for writing, editing, or formatting purposes and does \emph{not} impact the core methodology, scientific rigor, or originality of the research, declaration is not required.
    %this research? 
    \item[] Answer: \answerNA{} % Replace by \answerYes{}, \answerNo{}, or \answerNA{}.
    \item[] Justification: LLMs were not used as a core methodological component of this research.

\end{enumerate}

\end{document}